\documentclass[aps,eqsecnum,showpacs,amsmath,amssymb,preprint,floatfix]{revtex4}
\usepackage{amsmath}
\usepackage{bm}

\begin{document}

\title{Directional dependence of color superconducting gap in two-flavor QCD in a magnetic field}

\author{Lang Yu}
\email{langyu@asu.edu}
\affiliation{Department of Physics, Arizona State University, Tempe, Arizona 85287, USA}

\author{Igor A. Shovkovy}
\email{igor.shovkovy@asu.edu}
\affiliation{Department of Applied Sciences and Mathematics, Arizona State University, Mesa, Arizona 85212, USA}

\begin{abstract}
We study the effect of a magnetic field on the pairing dynamics in two-flavor color superconducting 
dense quark matter. The study is performed in the weakly coupled regime of QCD at asymptotically 
high density, using the framework of the Schwinger-Dyson equation in the improved rainbow approximation. 
We show that the superconducting gap function develops a directional dependence in momentum space. 
Quasiparticles with momenta perpendicular to the direction of the magnetic field have the largest gaps, 
while quasiparticles with momenta parallel to the field have the smallest gaps. We argue that the 
directional dependence is a consequence of a long-range interaction in QCD. The quantitative 
measure of the ellipticity of the gap function is determined by a dimensionless ratio, proportional to the 
square of the magnetic field and inversely proportional to the fourth power of the quark chemical potential. 
For magnetic fields in stars, $B\lesssim 10^{18}~\mbox{G}$, the corresponding ratio is estimated to be 
less than about  $10^{-2}$, justifying the use of the weak magnetic field limit in all stellar applications.
\end{abstract}

\date{February 3, 2012}


% 11.15.Ex	Spontaneous breaking of gauge symmetries
% 12.38.Aw	General properties of QCD (dynamics, confinement, etc.)
% 24.85.+p	Quarks, gluons, and QCD in nuclear reactions
% 26.60.-c	Nuclear matter aspects of neutron stars
 
\maketitle

\section{Introduction}
\label{Introduction}

Quantum chromodynamics (QCD), the fundamental theory of strong interactions, predicts that
quark matter at sufficiently high densities and sufficiently low temperatures is a color superconductor
\cite{Barrois1977, Bailin1982, sp1995, Alford1998, Rapp1998}. (For reviews, see for example 
Refs.~\cite{Shovkovy2005,Alford2008a}.) The property of asymptotic freedom in QCD ensures 
that such matter is weakly interacting at asymptotically large densities and, therefore, allows a 
rigorous treatment of the corresponding nonperturbative dynamics of Cooper pairing \cite{Son1999, 
Schafer1999, Shovkovy1999, Hong:1999ru, Hong2000, Hsu2000, Pisarski2000, self-e}. 
The simplest color superconducting phases correspond to spin-zero pairing. Depending on the 
number of quark flavors participating in pairing, one can have a two-flavor color superconducting 
(2SC) phase \cite{Alford1998, Rapp1998} or a color-flavor-locked (CFL) phase \cite{Alford1999}. 
Many additional complications arise when $\beta$-equilibrium and neutrality of quark 
matter is enforced \cite{dSC, g2SC, gCFL, instabilities}. 

Recently, the study of color superconductivity in the presence of magnetic fields attracted a lot of 
attention \cite{Ferrer2005, magCFL, Fayazbakhsh2010, Ferrer:2006ie, Feng:2009vt}. This interest 
is primarily driven by potential astrophysical applications, where magnetic fields play an important 
role. In the case of neutron stars, for example, the surface magnetic fields can reach up to about
$B\simeq 10^{12}~\mbox{G}$ \cite{reviewNS}. For magnetars, the corresponding fields can be still 
a few orders of magnitude larger, i.e., $B\simeq 10^{14}-10^{15}~\mbox{G}$, and perhaps even as 
high as $10^{16}~\mbox{G}$ \cite{Thompson1996}. Furthermore, it is possible that the magnetic fields  
in the stellar interiors are much higher and reach up to about $B\sim 10^{18}~\mbox{G}$ \cite{reviewNS,CPL}. 

In order to understand the properties of two- and three-flavor color superconducting phases with 
spin-zero pairing in a magnetic field, it is important to first recall their electromagnetic properties.
Despite being color superconductors, these phases can be penetrated by long-range 
``rotated'' magnetic fields, which are not subject to the Meissner effect \cite{Alford2000, 
Gorbar2000}. The rotated gauge fields are linear combinations of the vacuum photon 
and one of the gluons. While all Cooper pairs are neutral with respect to the corresponding 
rotated electromagnetism, the individual quark quasiparticles carry well defined charges.
It is not surprising, therefore, that the diquark pairing dynamics is affected by the presence 
of a magnetic field. The recent studies revealed many interesting qualitative features of the 
magnetic 2SC and CFL phases  \cite{Ferrer2005, magCFL, Fayazbakhsh2010}. However, 
all such studies share a common shortcoming: they are performed in the framework of 
Nambu-Jona-Lasinio (NJL) models with contact interactions. 

In this paper, we extend the analysis of two-flavor color superconductivity in a magnetic 
field by taking into account the long-range interaction in quark matter. In particular, we 
perform the study in the framework of the Schwinger-Dyson equation for the gap function 
in the weakly coupled regime of QCD at large densities. The long-range interaction is 
provided by the one-gluon exchange, in which the dominant screening and Landau 
damping effects are included. Our study reveals a qualitatively new feature of the 
magnetic 2SC phase, a directional dependence of the gap function, which is a 
consequence of the nonlocal interaction in quark matter. 

In the weak magnetic field limit, we find that the effect of a nonzero field can be mimicked 
by an effective increase of the strong coupling constant that governs the Cooper pairing 
dynamics: $g^2\to g^2 (1+\epsilon \sin^2\theta_{Bk})$, where $\theta_{Bk}$ is the angle between 
the quasiparticle momentum and the direction of the magnetic field, and the dimensionless 
quantity $\epsilon$ is a measure of ellipticity of the gap function. The latter is given by
the dimensionless ratio $\epsilon=27 \pi (eB)^2/(2 g^2 \bar{\mu}^4)$, where $B$ is the 
magnetic field and $\bar{\mu}$ is the quark chemical potential. As one can easily check,
this ratio is much less than $1$ even for the strongest possible fields in stars and, therefore, 
the use of the weak magnetic field limit is justified for all stellar applications. For completeness, 
we extend our analysis to the case of superstrong magnetic fields and find that the value 
of the gap increases with the field also in this regime. As expected on general grounds, 
the effects of nonlocality of the interaction become negligible in superstrong fields and 
the directional dependence of the gap disappears. It should be remarked, however, that our 
analysis  in the case of strong fields is performed with less rigor because the gluon screening 
effects in this case are not well known. 

The rest of the paper is organized as follows. In Sec.~\ref{Model}, we introduce the 
model and review how a constant rotated magnetic field enters the Lagrangian density
and how the rotated electric charges of quasiparticles are defined. Explicit expressions for 
quasiparticle propagators in sectors with different rotated charges are presented in 
Sec.~\ref{quark propagators}. Then, in Sec.~\ref{gap equation}, we derive the gap 
equation and solve it approximately in the limit of a weak magnetic field. In the same 
section, we also obtain an estimate for the gap in the strong field limit. In Sec.~\ref{conclusion}, 
we discuss the results and give a brief outlook. Several Appendices at the end 
of the paper contain many technical details and derivations used in the main text.

\section{Model}
\label{Model}

As stated in the introduction, the analysis in this study is done in the framework of weakly
interacting two-flavor QCD at large densities. The quadratic part of the corresponding 
Lagrangian density of quarks in an external rotated magnetic field is given by
\begin{eqnarray}\label{L^em}
  \mathcal
    {L}^{\rm em}_{\rm quarks} &=& \bar{\psi}\left(i\gamma^{\mu}\partial_{\mu}-m+\hat{\mu}\gamma^{0}
    +\tilde{e}\gamma^{\mu}\tilde{A}_{\mu}\tilde{Q}\right)\psi,
\end{eqnarray}
where $\tilde{A}_{\mu}$ is the rotated massless $\tilde{U}(1)_{\rm em}$ gauge field. This field is a linear 
combination of the vacuum photon ${A}_{\mu}$ and the eighth gluon $G^8_{\mu}$: 
$\tilde{A}_{\mu}=\cos\tilde\theta \, {A}_{\mu}-\sin\tilde\theta \, G^8_{\mu}$, where $\cos\tilde\theta = g/\sqrt{g^2+e^2/3}$
\cite{Alford2000,Gorbar2000}. (Here we use the standard convention for $SU(3)_c$ color generators 
in the adjoint representation \cite{Gorbar2000}.) The quarks carry flavor and color indices $\psi_{ia}$, 
where $i\in(u,d)=(1,2)$ is the flavor index and $a\in(r,g,b)=(1,2,3)$ is the color index. The multicomponent 
quark spinor field $\psi$ is assumed to have the following explicit form:
\begin{equation} 
\psi=\left(
\begin{array}{c}
\psi_{ur} \\[-2mm]
\psi_{ug} \\[-2mm]
\psi_{ub} \\[-2mm]
\psi_{dr} \\[-2mm]
\psi_{dg} \\[-2mm]
\psi_{db} 
\end{array}
\right).
\end{equation}
Here we assume that up and down quarks have the same masses 
($m_u=m_d=m$). In the 2SC phase, the matrix of chemical potentials $\hat{\mu}$ can have a nontrivial
color-flavor structure. When $\beta$-equilibrium and neutrality of quark matter is imposed \cite{g2SC}, 
the matrix elements of $\hat{\mu}$ read
\begin{equation} 
\label{chem-pots}
\mu_{ij, ab}= \left[\mu \, \delta_{ij}- \mu_e (Q_f)_{ij}\right] \delta_{ab} 
+ \frac{2}{\sqrt{3}}\mu_{8} \delta_{ij} (T_{8})_{ab},
\end{equation} 
where only one out of three parameters ($\mu$, $\mu_e$, and $\mu_8$) is truly independent, while 
the other two must be adjusted to achieve color and electric neutrality. For subtleties regarding the 
color neutrality; see Ref.~\cite{Buballa:2005bv}.

The explicit form of the quasiparticle charge operator $\tilde{Q}$, that corresponds to $\tilde{U}(1)_{\rm em}$ 
gauge group, is given by $\tilde{Q}=Q_f\otimes I_c-I_f\otimes (\frac{T_8}{\sqrt{3}})_c$, where 
$Q_f=\mbox{diag}(\frac{2}{3} ,-\frac{1}{3})$ is the usual matrix of electromagnetic charges of 
quarks in flavor space, and $T_8$ is the eighth generator of the $SU(3)_c$ gauge group in the adjoint 
representation. In units of $\tilde{e}=eg/\sqrt{g^2+e^2/3}$, the $\tilde{Q}$ charges of quarks are given in Table~\ref{charges}.
\begin{table}[ht]
\caption{$\tilde{Q}$ charges of quarks measured in units of $\tilde{e}=eg/\sqrt{g^2+e^2/3}$.}
\begin{ruledtabular}
\begin{tabular}{c c c c c c}
$u_r$ & $u_g$ & $u_b$ & $d_r$ & $d_g$ & $d_b$ \\
\hline
$+\frac{1}{2}$& $+\frac{1}{2}$  & 1 &$-\frac{1}{2}$ & $-\frac{1}{2}$ & 0 \\
\end{tabular}
\end{ruledtabular}
\label{charges}
\end{table}

In order to simplify the explicit form of the quark propagators in the magnetic 2SC phase, it is convenient 
to introduce the following set of projectors onto the subspaces of quasiparticles with different values of 
rotated-charges \cite{Fayazbakhsh2010}:
\begin{eqnarray}
  \Omega_{+\frac{1}{2}}=\mbox{diag}(1,1,0,0,0,0), \\
  \Omega_{+1}=\mbox{diag}(0,0,1,0,0,0),\\
  \Omega_{-\frac{1}{2}}=\mbox{diag}(0,0,0,1,1,0), \\
  \Omega_{0}=\mbox{diag}(0,0,0,0,0,1).
\end{eqnarray}
This is a complete set of projectors, satisfying the following relations:
\begin{eqnarray}
   &&\Omega_{\tilde{Q}}\Omega_{\tilde{Q}'} = \delta_{\tilde{Q}\tilde{Q}'}\Omega_{\tilde{Q}},\qquad 
   \tilde{Q},\tilde{Q}'=\pm1/2,+1,0, \\
  &&\Omega_{+\frac{1}{2}}+\Omega_{+1}+\Omega_{-\frac{1}{2}}+ \Omega_{0} = 1.
\end{eqnarray}
By making use of these projectors, we can decompose the multicomponent quark spinor field into 
separate pieces, describing groups of quasiparticles with different rotated charges:
\begin{equation}\label{spinor}
    \psi=\psi_{(+\frac{1}{2})}+\psi_{(+1)}+\psi_{(-\frac{1}{2})}+\psi_{(0)},
\end{equation}
where, by definition,
\begin{equation}\label{psi}
    \psi_{(+\frac{1}{2})}=\Omega_{+\frac{1}{2}}\psi,\qquad
    \psi_{(+1)}=\Omega_{+1}\psi,\qquad
    \psi_{(-\frac{1}{2})}=\Omega_{-\frac{1}{2}}\psi,\qquad
    \psi_{(0)}=\Omega_{0}\psi.
\end{equation}
In the new notation, the quadratic part of the quark Lagrangian density can be rewritten 
as follows:
\begin{eqnarray}
  \mathcal
    {L}^{\rm em}_{\rm quarks}  &=&\sum_{\tilde{Q}=\pm1/2,+1,0}
    \bar{\psi}_{(\tilde{Q})}(i\gamma^{\mu}\partial_{\mu}-m+{\mu_{\tilde{Q}}}\gamma^{0}
    +\tilde{e}\tilde{Q}\gamma^{\mu}\tilde{A}_{\mu})\psi_{(\tilde{Q})}.
    \label{L-sumQ}
\end{eqnarray}
As follows from Eq.~(\ref{chem-pots}), the chemical potentials $\mu_{\tilde{Q}}$ for quasiparticles
with different $\tilde{Q}$-charges, when projected onto the relevant color-flavor subspaces, are 
given by
\begin{eqnarray}
  \mu_{(+\frac{1}{2})}&=&\mu_{ur}=\mu_{ug}=\mu -\frac{2}{3}\mu_{e} +\frac{1}{3}\mu_{8}, \\
  \mu_{(-\frac{1}{2})}&=&\mu_{dr}=\mu_{dg} = \mu +\frac{1}{3}\mu_{e} +\frac{1}{3}\mu_{8}, \\
  \mu_{(+1)}&=&\mu_{ub}=\mu -\frac{2}{3}\mu_{e} -\frac{2}{3}\mu_{8}, \\
  \mu_{(0)}&=&\mu_{db}=\mu +\frac{1}{3}\mu_{e} -\frac{2}{3}\mu_{8}.
\end{eqnarray}
In this study, in order to simplify the analysis of the gap equation we will eventually neglect the 
effects due to nonzero $\mu_e$ and $\mu_8$. This is certainly justified in the study of QCD at 
asymptotically large densities. On the other hand, if the analysis is to be extrapolated to moderately large 
densities, relevant for compact stars, nonvanishing $\mu_e$ and $\mu_8$ may become 
important \cite{dSC, g2SC, gCFL, instabilities}. One should keep in mind, however, that the 
study of such a moderate density quark matter from first principles will be still quantitatively 
unreliable within the framework of the Schwinger-Dyson equation because of the strong coupling 
regime. As for the main purpose of this study, it aims only at a better understanding of the 
qualitative role of long-range forces.

\section{Quasiparticle propagators}
\label{quark propagators}

In the 2SC color superconducting phase, only the quasiparticles with the charges $\tilde{Q}=\pm\frac{1}{2}$
participate in Cooper pairing, while the remaining two quasiparticles (with charges $\tilde{Q}=0,1$) play the 
role of passive spectators. Therefore, in the rest of the analysis, we will concentrate exclusively on the two 
pairs of quasiparticles participating in Cooper pairing and ignore the others. 

As usual in studies of color superconducting phases, it is convenient to introduce the 
Nambu-Gorkov spinors,
\begin{equation}\label{NG}
   \bar{\Psi}_{(\tilde{Q})}=(\bar{\psi}_{(\tilde{Q})},\bar{\psi}_{(-\tilde{Q})}^{C}), \qquad  
   \Psi_{(\tilde{Q})}=\left(
                          \begin{array}{c}
                            \psi_{(\tilde{Q})} \\
                            \psi^C_{(-\tilde{Q})} \\
                          \end{array}
                        \right),
\end{equation}
for quasiparticles with the charges $\tilde{Q}=\pm\frac{1}{2}$. 
Here $\psi_{(\tilde{Q})} ^C=C\bar{\psi}_{(\tilde{Q})}^T$ 
and $\bar{\psi}_{(\tilde{Q})}^{C}={\psi}_{(\tilde{Q})}^TC$ 
are the charge-conjugate spinors, 
and $C=i\gamma^2\gamma^0$ is the charge-conjugation matrix satisfying the relations:  
$C^{-1}\gamma^\mu C=-(\gamma^\mu)^T$ and $C=-C^T$. 
In terms of the Nambu-Gorkov spinors, Lagrangian density (\ref{L-sumQ}) takes the form
\begin{equation}\label{L}
    \mathcal
    {L}^{\rm em}_{\rm quarks} =\frac{1}{2}\sum_{\tilde{Q}=\pm1/2} \bar{\Psi}_{(\tilde{Q})}S^{-1}_{(\tilde{Q}),0}\Psi_{(\tilde{Q})}
    +\sum_{\tilde{Q}=+1,0} \bar{\psi}_{(\tilde{Q})}[G^+_{(\tilde{Q}),0}]^{-1}\psi_{(\tilde{Q})},
\end{equation}
where the inverse free propagator $S^{-1}_{(\tilde{Q}),0}$ for each sector with a fixed value of $\tilde{Q}$-charge 
has a block-diagonal form,
\begin{equation}\label{S0}
    S_{(\tilde{Q}),0}^{-1}=\mbox{diag}\left(   [G^+_{(\tilde{Q}),0}]^{-1} , [G^-_{(\tilde{Q}),0}]^{-1}  \right),
\end{equation}
and the explicit form of the diagonal elements reads
\begin{equation}
\left[G^{\pm}_{(\tilde{Q}),0}\right]^{-1} = \gamma^{\mu}\left(i\partial_{\mu} + \tilde{Q}\tilde{e}\tilde{A}_{\mu}\right)
  \pm \mu_{(\tilde{Q})}\gamma^0-m.
\end{equation}

For quasiparticles participating in Cooper pairing, the full propagators also have nonzero off-diagonal 
Nambu-Gorkov components, determined by the color superconducting gap function, i.e.,
\begin{equation}\label{S(K)}
    S^{-1}_{(\tilde{Q})}=\left(
                    \begin{array}{cc}
                      [G^+_{(\tilde{Q}),0}]^{-1} & \Delta^-_{(\tilde{Q})} \\
                      \Delta^+_{(\tilde{Q})} & [G^-_{(\tilde{Q}),0}]^{-1} \\
                    \end{array}
                  \right).
\end{equation}
The color-flavor structures of $\Delta^-_{(\tilde{Q})}$ and $\Delta^+_{(\tilde{Q})}$ are given by
\begin{eqnarray}\label{Delta-}
    \Delta^-_{(+\frac{1}{2})}&=&-\Delta^-_{(-\frac{1}{2})}=\left(
                                \begin{array}{cc}
                                  0 & -i\gamma^5\Delta \\
                                  i\gamma^5\Delta & 0 \\
                                \end{array}
                              \right), \\
    \Delta^+_{(+\frac{1}{2})} &=& -\Delta^+_{(-\frac{1}{2})}=\left(
                                \begin{array}{cc}
                                  0 & i\gamma^5\Delta^*  \\
                                  -i\gamma^5\Delta^* & 0  \\
                                \end{array}
                              \right).     
\end{eqnarray}
Note that the explicit forms of the two relevant Nambu-Gorkov spinors (\ref{NG}) read
\begin{equation}
\label{Psi+1/2 Psi-1/2}
    \Psi_{(+\frac{1}{2})}=\left(
                           \begin{array}{c}
                             \psi_{ur} \\[-1mm]
                             \psi_{ug} \\[-1mm]
                             \psi_{dr}^C \\[-1mm]
                             \psi_{dg}^C 
                           \end{array}
                         \right),
\qquad
    \Psi_{(-\frac{1}{2})}=\left(
                           \begin{array}{c}
                             \psi_{dr} \\[-1mm]
                             \psi_{dg} \\[-1mm]
                             \psi_{ur}^C \\[-1mm]
                             \psi_{ug}^C 
                           \end{array}
                         \right).
\end{equation}
It appears that one can partially diagonalize the inverse full propagators $S_{(\tilde{Q})}^{-1}$ by simply 
reordering the components of the spinors as follows:
\begin{equation}
\label{Psi+1/2 Psi-1/2 new}
    \Psi^{\rm new}_{(+\frac{1}{2})}=\left(
                           \begin{array}{c}
                            \psi_{ur} \\[-1mm]
                            \psi_{dg}^C \\[-1mm]
                            \psi_{ug} \\[-1mm]
                            \psi_{dr}^C 
                           \end{array}
                         \right),
\qquad
    \Psi^{\rm new}_{(-\frac{1}{2})}=\left(
                           \begin{array}{c}
                             \psi_{dr} \\[-1mm]
                             \psi_{ug}^C \\[-1mm]
                             \psi_{dg} \\[-1mm]
                             \psi_{ur}^C 
                           \end{array}
                         \right).
\end{equation}
From physics viewpoint, the possibility of such a partial diagonalization reflects the fact that there are 
two different types of Cooper pairs: one made of red up and green down quarks and the other made 
of green up and red down quarks.  

In the new basis, the inverse full propagator $S_{(\tilde{Q})}^{-1}$ has the following block-diagonal form:
\begin{equation}
\label{inv-fullS-Q}
S_{(\tilde{Q})}^{-1} = \mbox{diag} \left( [S_{(\tilde{Q})}^X]^{-1} , [S_{(\tilde{Q})}^Y]^{-1}  \right),
\end{equation}
where 
\begin{equation}\label{diagS1/2^1}
    [S_{(+\frac{1}{2})}^X]^{-1}=\left(
                                     \begin{array}{cc}
                                       \gamma^{\mu}(i\partial_{\mu}+ \frac{1}{2}\tilde{e}\tilde{A}_{\mu})+\mu_{ur}\gamma^0-m & -i\gamma^5\Delta \\
                                       -i\gamma^5\Delta^* & \gamma^{\mu}(i\partial_{\mu}+ \frac{1}{2}\tilde{e}\tilde{A}_{\mu})-\mu_{dg}\gamma^0-m \\
                                     \end{array}
                                   \right),
\end{equation}
\begin{equation}\label{diagS1/2^2}
    [S_{(+\frac{1}{2})}^Y]^{-1}=\left(
                                     \begin{array}{cc}
                                       \gamma^{\mu}(i\partial_{\mu}+ \frac{1}{2}\tilde{e}\tilde{A}_{\mu})+\mu_{ug}\gamma^0-m & i\gamma^5\Delta \\
                                       i\gamma^5\Delta^* & \gamma^{\mu}(i\partial_{\mu}+ \frac{1}{2}\tilde{e}\tilde{A}_{\mu})-\mu_{dr}\gamma^0-m \\
                                     \end{array}
                                   \right),
\end{equation}
and
\begin{equation}\label{diagS-1/2^1}
    [S_{(-\frac{1}{2})}^X]^{-1}=\left(
                                     \begin{array}{cc}
                                       \gamma^{\mu}(i\partial_{\mu}- \frac{1}{2}\tilde{e}\tilde{A}_{\mu})+\mu_{dr}\gamma^0-m & i\gamma^5\Delta \\
                                       i\gamma^5\Delta^* & \gamma^{\mu}(i\partial_{\mu}- \frac{1}{2}\tilde{e}\tilde{A}_{\mu})-\mu_{ug}\gamma^0-m \\
                                     \end{array}
                                   \right),
\end{equation}
\begin{equation}\label{diagS-1/2^2}
    [S_{(-\frac{1}{2})}^Y]^{-1}=\left(
                                     \begin{array}{cc}
                                       \gamma^{\mu}(i\partial_{\mu}- \frac{1}{2}\tilde{e}\tilde{A}_{\mu})+\mu_{dg}\gamma^0-m & -i\gamma^5\Delta \\
                                       -i\gamma^5\Delta^* & \gamma^{\mu}(i\partial_{\mu}- \frac{1}{2}\tilde{e}\tilde{A}_{\mu})-\mu_{ur}\gamma^0-m \\
                                     \end{array}
                                   \right).
\end{equation}
Using the representation for the inverse quasiparticle propagator in Eq.~(\ref{inv-fullS-Q}), we find the propagator itself, 
\begin{equation}
\label{fullS-Q new}
S_{(\tilde{Q})} = \mbox{diag} \left( S_{(\tilde{Q})}^X  , S_{(\tilde{Q})}^Y \right).
\end{equation}
The calculation of the corresponding diagonal blocks $S_{(\tilde{Q})}^{X,Y}$ is tedious, but straightforward. 
The details of derivation are presented in Appendix~\ref{Appendix A}.

\section{Gap equation} 
\label{gap equation}

In the coordinate space, the gap equation (i.e., the off-diagonal component of the Schwinger-Dyson
equation for the full propagator) reads
\begin{equation}\label{gap}
     \left[S^X_{(\tilde{Q})}\right]^{-1}_{21}(u,u^\prime)=i g^2\gamma^{\mu} \left(-T^{A}\right)^T
     \left[S^X_{(\tilde{Q})}\right]_{21}(u,u^\prime)\gamma^{\nu} T^{B} D_{\mu\nu}^{AB}(u,u^\prime),
\end{equation}
where $D_{\mu\nu}(u,u^\prime)$ is the gluon propagator, and $u\equiv (t,z,\bm{r}_{\perp})$ is a four-vector 
of space-time position. We will assume that the gluon propagator is diagonal in adjoint color indices. 
Note that the off-diagonal component of the propagator $S^Y_{(\tilde{Q})}$ satisfies a similar equation. 
While Eq.~(\ref{gap}) describes Cooper pairing of red up and green down quarks, the equation for 
$S^Y_{(\tilde{Q})}$ describes Cooper pairing of green up and red down quarks. 

In this study of Cooper pairing in a magnetized color superconducting phase, it is convenient to 
start from the coordinate-space representation of the gap equation [see Eq.~(\ref{gap})] and 
then switch to the Landau-level representation. This is in contrast to the usual momentum 
space representation, often utilized in the case of vanishing external fields. 

In this connection, a short remark is in order regarding the general structure of a quasiparticle 
propagator. Because of the interaction of charged quasiparticles with the magnetic field, their 
momenta in the two spatial directions perpendicular to the field are not well-defined quantum 
numbers. This is reflected in the structure of the propagator (as well as its inverse), which is not a 
translationally invariant function in coordinate space. Instead, the quasiparticle propagator 
has the form of a product of the universal Schwinger phase (which spoils the translational 
invariance) and a translationally invariant part \cite{schwinger} (for details, see 
Appendix~\ref{Appendix A}). 

After factoring out the same Schwinger phase on both sides of the gap equation and 
projecting the resulting equation onto subspaces of different Landau levels, one obtains 
an infinite set of coupled equations; see Eq.~(\ref{Full gap A}) in Appendix~\ref{Appendix B}. 
For both charges $\tilde{Q}=\pm 1/2$, the gap equations are similar. Here we show only the 
final set of equations for $\tilde{Q}=+1/2$:
\begin{eqnarray}\label{Full gap}
\Delta_m\mathcal{P}_{-}+\Delta_{m+1}\mathcal{P}_{+} & =&
  -i\frac{2g^2}{3}\sum_{n=0}^{\infty}\int\frac{d\omega' dk'^3}{(2\pi)^2}
   \int \frac{d^2\bm{q}_{\perp}}{(2\pi)^2}\gamma^{\mu} \Delta_n \bigg[ 
   \mathcal{L}^{(0)}_{n,m} \frac{\mathcal{E}_n}{\mathcal{C} _n}\mathcal{P}_{-}  +\mathcal{L}^{(0)}_{n-1,m}
  \frac{\mathcal{E}_n}{\mathcal{C} _n}\mathcal{P}_{+} \bigg]\nonumber\\
  &\times&
  \gamma^{\nu}
  D_{\mu\nu}(\omega-\omega',k^3-k'^3;\bm{q}_{\perp}),
\end{eqnarray}
where $m,n=0,1,2,\ldots$ are Landau-level indices, and functions $\mathcal{C}_n$ and $\mathcal{E}_n$ 
are defined in Appendix~\ref{Appendix A}; see Eq.~(\ref{C-n}) and (\ref{E-n}), respectively. These functions 
depend on the parameters of the model (e.g., masses and chemical potentials of quarks) as well as on the 
color superconducting gap parameters $\Delta_{n}$. Note that the gaps associated with different Landau 
levels are not necessarily equal. This fact is emphasized by the Landau-level subscript $n$ in the notation.
Here and below, we assume that all gaps $\Delta_{n}$ are real functions.

\subsection{Gluon propagator}
\label{Gluon propagator}

In dense quark matter, unlike in vacuum, the gluon exchange interaction is partially screened. 
Therefore, when analyzing the Cooper pairing dynamics between quarks, it is very important 
to take the relevant screening effects due to nonzero density into consideration \cite{Son1999}. 
In the problem at hand, in addition, one should account for the external magnetic field, which 
can further modify the screening of the one-gluon interaction through quark loops. The latter 
can be quite important in strong magnetic fields \cite{Miransky:2002rp}. To simplify the analysis 
in this study, we will assume that the magnetic field is weak ($|eB|\ll \mu^2$). At the end, we shall 
see that this happens to be a very good approximation for most stellar applications. 

In the case of a weak external field, the screening of the one-gluon interaction in dense 
medium can be described well by the usual hard-dense loop approximation \cite{Heinz:1985qe,
Vija:1994is,Manuel:1995td}. In the Coulomb gauge, the Lorentz structure of the gluon propagator 
is given by \cite{Pisarski1989,LeBellac1996}
\begin{equation}\label{gluon M}
 D_{\mu\nu}(Q)=-\frac{Q^2}{q^2}\frac{\delta_{\mu0}\delta_{\nu0}}{Q^2-F}
 -\frac{P^T_{\mu\nu}}{Q^2-G},
\end{equation}
where functions $F$ and $G$ define the spectra of the longitudinal and transverse gluons, 
respectively. Both functions depend on the energy $q^0$ and the absolute value of the 
three-momentum $|\vec{q}|$. By definition, $Q=(q^0,\vec{q})$ is a momentum four-vector. 
The transverse Lorentz projector $P^T_{\mu\nu}$ is defined as follows:
\begin{equation} 
 P^T_{00} = P^T_{0i}=0,\qquad 
 P^T_{ij} =\delta_{ij}-\hat{q}_i\hat{q}_j.
\end{equation}
In the most important regime for Cooper pairing dynamics, $q^0\ll |\vec{q}|\ll m_D$, the
approximate expressions for these screening functions read \cite{Heinz:1985qe,Vija:1994is,Manuel:1995td}
\begin{equation}\label{app}
   F\simeq m_D^2, \qquad 
   G\simeq \frac{\pi}{4}m_D^2\frac{q^0}{|\vec{q}|},
\end{equation}
where $m_D^2=(g\mu/\pi)^2$ is the Debye screening mass in two-flavor quark matter. At large densities, 
the exchange interaction by electric gluon modes is strongly suppressed due to Debye screening 
and, to leading order, plays no role. Magnetic gluon modes, on the other hand, are subject only to 
a mild dynamical screening (Landau damping) at nonzero frequencies and play the dominant role 
in Cooper pairing \cite{Son1999}.

\subsection{Gap equation: Weak magnetic field limit}
\label{the gap equation in the weak magnetic field limit}

In order to obtain the gap equation in the weak magnetic field limit, we expand the
translationally invariant part of the full fermion propagator in powers of the magnetic field and 
keep the leading terms up to second order, $(\tilde{e}\tilde{Q}\tilde{B})^2$ (for details, 
see Appendixes~\ref{Appendix C} and \ref{Appendix D}). Omitting the technical details, 
here we present the final form of the gap equation,
\begin{equation}\label{gap Eq weak}
  \Delta(\omega) =T^{(0)}(\omega)+T^{(1)}(\omega)+T^{(2)}(\omega),
  \end{equation}
where 
  \begin{equation} 
    T^{(i)}(\omega)= -i\frac{2g^2}{3}\int\frac{d\omega^\prime}{2\pi}
   \int \frac{d^3\bm{k}^\prime}{(2\pi)^3}   \Delta(\omega')  \gamma^{\mu} K^{(i)}(\omega^\prime,\bm{k}^\prime)
   \gamma^{\nu}
   D_{\mu\nu}(\omega-\omega^\prime,\bm{k}-\bm{k}^\prime)
   \label{Tis}
\end{equation}
is the contribution of the $i$th order in powers of the magnetic field. The explicit form of the kernels 
$K^{(i)}(\omega,\bm{k})$ for the three leading-order terms in the gap equation are presented in 
Eqs.~(\ref{K^(0)}), (\ref{K^(1)}) and (\ref{K^(2)}) in Appendix~\ref{Appendix C}.

At zero magnetic field, Eq.~(\ref{gap Eq weak}) reduces to the well-known gap equation in the 2SC 
phase without a magnetic field \cite{Son1999, Schafer1999, Shovkovy1999, Hong:1999ru, Hong2000, 
Hsu2000, Pisarski2000}. After switching to the Euclidean space and performing the traces on both 
sides of the gap equation, we rederive the following zeroth-order (i.e., vanishing magnetic field) 
equation:
\begin{eqnarray}\label{gap-eq-0th-order}
   \Delta^{(0)}(\omega_E) &=&  \frac{g^2}{3}\int\frac{d\omega_E^\prime }{(2\pi)}
   \int \frac{d^3\bm{k}^\prime}{(2\pi)^3}  \frac{\Delta^{(0)}(\omega_E^\prime)}
   {(\omega_E^\prime)^2+({k^\prime}-\bar{\mu})^2+[\Delta^{(0)}(\omega_E^\prime)]^2}\nonumber \\
    &&\times\bigg[\frac{1}{(\omega_E-\omega_E^\prime)^2+|\bm{k}-\bm{k}^\prime|^2+m_D^2}
    +\frac{2|\bm{k}-\bm{k}^\prime|}{|\bm{k}-\bm{k}^\prime|^3+\omega_l^3}\bigg],
\end{eqnarray}
where $ k=|\bm{k}|$, $ k^\prime=|\bm{k}^\prime|$, $\omega_E=i\omega$, $\omega_E^\prime=i\omega^\prime$, and 
$\omega_l^3=(\pi/4)m_D^2|\omega_E^\prime-\omega_E|$. For simplicity, here we assumed that $m=0$ and that 
the chemical potentials of all quarks are identical and equal $\bar{\mu}$. 

After performing the integration over $\bm{k}^\prime$ and keeping only the leading-order 
contributions from the dynamically screened magnetic gluon exchange, we arrive at
 \begin{eqnarray}\label{gap0}
  \Delta^{(0)}(\omega_E)
    &=&\frac{g^2}{36\pi^2}\int_{-\infty}^{\infty}{d\omega_E^\prime}
    \frac{\Delta^{(0)}(\omega_E^\prime)}{\sqrt{(\omega_E^\prime)^2+(\Delta^{(0)})^2}}
    \ln\frac{\Lambda}{|\omega_E'-\omega_E|},
\end{eqnarray}
where $\Lambda={4(2\mu)^3}/({\pi m_D^2})$. The approximate solution to this equation reads
\cite{Son1999, Schafer1999, Shovkovy1999, Hong:1999ru, Hong2000, Hsu2000, Pisarski2000}
\begin{equation}
  \Delta^{(0)} \simeq \Lambda \exp(-\frac{3\pi^2}{\sqrt{2}g}+1).
  \label{gap-b0}
\end{equation}
Using this result as a benchmark, let us proceed to the case of a weak but nonzero magnetic field. 

It is easy to check (and might have been expected from the symmetry arguments) that the first-order term, 
i.e., $T^{(1)}(\omega)$ in Eq.~(\ref{Tis}), which is linear in a magnetic field, vanishes after the Dirac traces 
are performed. Thus, the leading correction to the gap equation in a weak magnetic field comes from the 
second-order term, i.e., $T^{(2)}(\omega)$ in Eq.~(\ref{Tis}). 

To the same leading order in coupling, which includes only the exchange interaction due to dynamically 
screened magnetic gluons, we derive the following explicit form of the gap equation (for the details of derivation, 
see Appendix~\ref{Appendix D}):
\begin{eqnarray}
  \Delta^{(B)}(\omega_E)
    &=& \frac{g^2}{36\pi^2}  \int_{-\infty}^{\infty} d\omega_E^\prime  
   \Delta^{(B)}(\omega_E^\prime)\Bigg[\frac{1}{\sqrt{(\omega_E^\prime)^2+(\Delta^{(B)})^2}}
   \ln\frac{\Lambda}{|\omega_E^\prime-\omega_E|}+\nonumber\\
  &&+\frac{9\omega_l^{15} (\tilde{e}\tilde{Q}\tilde{B})^2  \sin^2\theta_{Bk} }{4\bar{\mu}^2\left(\omega_l^6+\left[(\omega_E^\prime)^2+(\Delta^{(B)})^2\right]^3\right)^3}\ln\frac{\omega_l}{|\omega_E^\prime-\omega_E|}\Bigg].
  \label{gap-eq-final}
\end{eqnarray}
The detailed analysis of this equation may not be very easy. However, several of its qualitative properties  
are obvious right away. First of all, the positive sign of the subleading-order correction, proportional to 
$(\tilde{e}\tilde{Q}\tilde{B})^2$, indicates that the gap increases with the magnetic field. This is in 
qualitative agreement with the intuitive expectation that the external magnetic field should enhance 
the binding energy of Cooper pairs made of quasiparticles with opposite charges \cite{Ferrer2005,
magCFL,Fayazbakhsh2010}. From the fact that this correction to the gap equation is also proportional 
to $\sin^2\theta_{Bk}$, where $\theta_{Bk}$ is the angle between the quasiparticle momentum and 
the magnetic field, we conclude that the gap function acquires a directional dependence. Moreover,
we see that the largest value of the gap will be for quasiparticles with the momenta perpendicular
to the magnetic field. On the other hand, for quasiparticles with the momenta parallel to the field, 
there is no enhancement of the gap at all. 

In order to understand the qualitative effect of the subleading term quadratic in magnetic field,
we can perform the following semirigorous analysis of Eq.~(\ref{gap-eq-final}). To this end, let 
us cut the infrared region of integration off at $\omega_E^\prime = \Delta^{(B)}$ and substitute 
$\Delta^{(B)}=0$ in the denominators of both terms on the right-hand side of the equation. We 
then arrive at
\begin{equation}
\Delta^{(B)}  \simeq \frac{g^2}{18\pi^2} \left(1+\frac{54 \pi (\tilde{e}\tilde{Q}\tilde{B})^2 }{g^2 \bar{\mu}^4} \sin^2\theta_{Bk} \right)
 \int_{\Delta^{(B)}}^{\Lambda} d\omega_E^\prime  \frac{\Delta^{(B)}  }{\omega_E^\prime}\ln\frac{\Lambda}{\omega_E^\prime}.
\end{equation}
While this approximation cannot be used to get a reliable estimate for the gap, it is very helpful 
to understand the qualitative effect of the magnetic field on the pairing dynamics in color superconducting
dense quark matter. It shows that the effective coupling constant in the presence of a magnetic field becomes 
larger, i.e.,
\begin{equation}
g^2\to g_{\rm eff}^2= g^2\left(1+\frac{27\pi (\tilde{e}\tilde{B})^2 }{2g^2 \bar{\mu}^4} \sin^2\theta_{Bk} \right),
\label{g-eff}
\end{equation}
where we substituted $\tilde{Q}=\pm\frac{1}{2}$. The validity of the weak field approximation requires 
that the subleading correction is small compared to the leading result. This translates into the requirement 
$|\tilde{e}\tilde{B}|^2 \lesssim g^2 \bar{\mu}^4$. As we shall see below, this condition is always 
satisfied in stellar applications. 

Without rigorously solving the gap equation (\ref{gap-eq-final}), now we can claim that the solution for 
the gap function in the magnetic 2SC phase is approximately given by the same expression as in the 
absence of the field, but with the coupling constant $g$ replaced by $g_{\rm eff}$, i.e.,
\begin{equation}\label{gap-result}
  \Delta^{(B)} \simeq \Lambda \exp(-\frac{3\pi^2}{\sqrt{2}g_{\rm eff}}+1)
  \simeq \Delta^{(0)} e^{\beta_{Bk}},
\end{equation}
where the explicit expression for $\beta_{Bk}$ follows from Eq.~(\ref{g-eff}), 
\begin{equation}
\beta_{Bk}=\frac{81\pi^3 (\tilde{e}\tilde{B})^2}{4\sqrt{2}g^3\bar{\mu}^4} \sin^2\theta_{Bk}.
\end{equation}
This is a nonnegative function, which depends on the angle between the quasiparticle momentum 
$\bm{k}$ and the magnetic field $\tilde{\bm{B}}$. Its maximum value $\beta_{Bk}^{\rm (max)}$ is 
obtained at $\theta_{Bk}= 90^{\circ}$.

The final result in Eq.~(\ref{gap-result}) is interesting for several reasons. Most importantly, it shows that 
the gap is nonisotropic, taking its largest values when the quasiparticle momenta are perpendicular to 
the direction of the magnetic field, and taking its smallest value when the quasiparticle momenta are 
along/against the field. We also find that, compared to the case without the magnetic field, the gap is 
subject to an increase in all directions of quasiparticle momenta, except for the directions exactly along
or against the magnetic field.

\subsection{Gap equation: Strong magnetic field limit}
\label{the gap equation in the strong magnetic field limit}

To get a qualitative insight about the pairing dynamics in the case of a strong magnetic field, 
$\tilde{e}\tilde{B}\gtrsim \mu^2$, it seems sufficient to consider the gap equation in the lowest 
Landau-level approximation. The choice of a simple approximation for the gluon exchange 
interaction is much harder to justify. Here we will use the gluon propagator with the screening 
effects at zero magnetic field. Obviously, such an approximation is not very reliable. A naive 
justification for such an approximation is the observation that gluons couple not only to 
the charged quasiparticles (with $\tilde{Q}=\pm 1/2$ and $\tilde{Q}=1$), which are strongly 
affected by the magnetic field, but also to neutral quasiparticles (with $\tilde{Q}=0$), which 
are not affected by the magnetic field at all. If zero density ($\mu=0$) and strong magnetic 
field limit in gauge theories is used as a guide for intuition, one may suggest that those gluons, 
which are coupled only to charged quasiparticles, will be subject to an additional Debye 
screening with an effective mass $m_{D}^{\rm eff} \propto g \sqrt{|\tilde{e} \tilde{B}|}$ 
\cite{Miransky:2002rp}. The other gluons will be still providing the same dominant 
interaction with dynamical screening as in absence of the external field. Then, the 
usual hard dense loop approximation may be still qualitatively reasonable. Besides, 
to the best of our knowledge, the explicit result for the polarization tensor (screening) 
in dense QCD matter ($\mu\neq 0$) in a magnetic field ($B\neq 0$) is not available 
in the literature. Thus, the main purpose of our exercise in this subsection, which is
based on the simplest possible approximation, will be to roughly estimate the color 
superconducting  gap due to long-range interaction in the regime of a strong external 
magnetic field. 

By making use of Eq.~(\ref{Full gap}), we easily derive the gap equation in the lowest 
Landau-level  approximation,
\begin{eqnarray}
   \Delta^{(B)}(\omega_E) &=&  \frac{g^2}{3}\int\frac{d\omega_E^\prime dk'^3}{(2\pi)^2}
   \int \frac{d^2\bm{q}_{\perp}}{(2\pi)^2} \exp\left(-\frac{{q}_{\perp}^2l^2}{2}\right)
   \frac{\Delta^{(B)}(\omega_E^\prime) }{(\omega_E^\prime)^2+(k'^3-{\mu})^2+(\Delta^{(B)})^2}\nonumber \\
  &\times&  \frac{{q}_{\perp}^2}{(k'^3-k^3)^2+{q}_{\perp}^2} 
   \frac{[(k'^3-k^3)^2+{q}_{\perp}^2]^{\frac{1}{2}}}{[(k'^3-k^3)^2+{q}_{\perp}^2]^{\frac{3}{2}}+\omega_l^3}.
\end{eqnarray}
Because of the exponential suppression in the integration over the transverse momentum $q_{\perp}$, 
the dominant contribution comes from the region of small momenta, $q_{\perp} l \lesssim 1$. Therefore, 
an approximate result can be obtained by simply making a sharp ultraviolet cutoff at $q_{\perp}=\sqrt{2}/l$ 
and dropping altogether the exponential factor $\exp(- {q}_{\perp}^2l^2/2)$ in the integrand. After performing 
the integration also over the longitudinal momentum $k'^3$, we will arrive at the following approximate 
gap equation:
\begin{equation}\label{final gap Eq strong}
  \Delta^{(B)}(\omega_E)
   \approx \frac{g^2}{72\pi^2}\int^{+\infty}_{-\infty}d\omega_E^\prime  
   \frac{\Delta^{(B)}(\omega_E^\prime) }{\sqrt{(\omega_E^\prime)^2+(\Delta^{(B)})^2}}
   \ln\frac{\Lambda_B}{|\omega_E^\prime-\omega_E|},
\end{equation}
where $\Lambda_B=\frac{8\sqrt{2}}{\pi m_D^2 l^3}=\frac{8\pi\sqrt{2}|\tilde{e}\tilde{Q}\tilde{B}|^{3/2}}{g^2 \bar{\mu}^2}$.
As we see, this equation has the same structure as Eq.~(\ref{gap0}), but with a smaller effective coupling and a 
different expression for $\Lambda_B$. Making use of this fact, we can get an approximate solution for the gap in the limit 
of strong magnetic field by properly modifying the result in Eq.~(\ref{gap-b0}), i.e.,
\begin{equation}\label{gap strong solution}
\Delta^{(B)} =\frac{4 \pi |\tilde{e}\tilde{B}|^{3/2}}{g^2 \bar{\mu}^2} \exp(-\frac{3\pi^2}{g}+1).
\end{equation}
Here we substituted $\tilde{Q}=\pm\frac{1}{2}$. This result shows that the strong magnetic field  
strengthens the diquark pair formation.  This is in qualitative agreement with the findings in models with
local interaction \cite{Ferrer2005,magCFL,Fayazbakhsh2010}. 

In contrast to the result in the weak magnetic field limit, there is no directional dependence in the gap 
function when the field is strong. This suggests that the corresponding pairing dynamics is essentially 
local. While the result may appear surprising at first sight, this finding in fact agrees with the intuitive 
picture that the motion of charged particles is restricted over distances of the order of the magnetic 
length, $l=1/\sqrt{|\tilde{e}\tilde{Q}\tilde{B}|}$, in the plane perpendicular to the magnetic field. When 
Cooper pairs form, the additional spatial restriction on particles' motion (partial localization) can 
strongly enhance the binding energy and substantially reduce the size of bound states.

\section{Conclusion} 
\label{conclusion}
non
In this paper, we studied the effect of a rotated magnetic field on the Cooper pairing dynamics in the
two-flavor color superconducting phase of dense quark matter with long-range interaction provided by 
the one-gluon exchange with dynamical screening. Using the Landau-level representation, we derived 
a set of gap equations valid for an arbitrary magnetic field. These equations show that, in general, 
the gaps are functions of the Landau-level index $n$. Therefore, solving the corresponding set of 
equations may be rather involved and require the use of sophisticated numerical methods. Instead,
here we used analytical methods to investigate the limiting cases of weak and strong magnetic 
fields. 

In the weak magnetic field limit, the energy separation between the Landau levels is vanishingly small
and there is no reason to expect a strong dependence of the gaps on the corresponding discrete index 
$n$. This justifies the use of an approximation in which the gaps are the same in all Landau levels 
near the Fermi surface. Additionally, in this case the quasiparticle propagator allows a simple expansion 
in powers of the magnetic field that greatly simplifies the structure of the resulting gap equation; 
see Eq.~(\ref{gap Eq weak}). We find that the leading-order term, affecting the gap, is quadratic 
in the magnetic field. The corresponding correction to the vanishing magnetic field result for the 
gap is determined by the value of parameter 
$\beta_{Bk}^{{\rm (max)}}=81\pi^3 (\tilde{e}\tilde{B})^2/(4\sqrt{2} g^3\bar{\mu}^4)$, where $\tilde{B}$ 
is the magnetic field and $\bar{\mu}$ is the quark chemical potential; see Eq.~(\ref{gap-result}). 
The numerical value of this parameter appears to be quite small even for strongest possible 
magnetic fields in compact stars, $\tilde{B} \lesssim 10^{18}~\mbox{G}$. Indeed, the 
corresponding numerical estimate reads
\begin{equation}
\beta_{Bk}^{{\rm (max)}} \approx 1.3 \times 10^{-2} \left(\frac{400~\mbox{MeV}}{\bar{\mu}}\right)^4 
\left(\frac{\tilde{B}}{10^{18}~\mbox{G}}\right)^2.
\end{equation}
(Here, for the strong coupling constant, we used $g= \sqrt{4\pi}$, which corresponds to $\alpha_{s} =1$.)

The most interesting feature of the pairing dynamics in the presence of a magnetic field is a directional 
dependence of the gap function in momentum space. The magnetic field correction to the gap is 
proportional to $ \sin^2\theta_{Bk}$, where $\theta_{Bk}$ is the angle between the quasiparticle 
momentum $\bm{k}$ and the magnetic field $\tilde{\bm{B}}$. From the physics viewpoint, this means that quasiparticles 
with momenta pointing perpendicular to the direction of the magnetic field have the largest gaps, while 
quasiparticles with momenta along/against the field have the smallest gaps. Clearly, such a directional 
dependence is a qualitative outcome of a long-range interaction in the model used. This contrasts 
with the studies based on models with pointlike interactions in Refs.~\cite{Ferrer2005, magCFL, 
Fayazbakhsh2010}, where the gaps are always isotropic.

Our analysis in the case of a strong magnetic field is admittedly less rigorous. We use the lowest 
Landau-level approximation and utilize the simplest approximation for the gluon exchange interaction 
without modifying the screening effects due to a nonzero magnetic field. The resulting estimate for the gap 
is given in Eq.~(\ref{gap strong solution}). Our result shows that strong magnetic fields enhance the 
diquark Cooper pairing and lead to larger color superconducting gaps. This is in qualitative agreement 
with the findings in Refs.~\cite{Ferrer2005, magCFL, Fayazbakhsh2010}, where the models 
with short-range interactions were used. We also find that, because of the partial localization of 
quasiparticles in a strong magnetic field, the corresponding dynamics is essentially local and there 
is no directional dependence of the gap. 

To go beyond the two limiting cases, analyzed in this paper, one will need to properly truncate an 
infinite set of gap equations and use numerical methods to solve it. In such an approach, it may be 
also possible to include the effects of different quark masses and chemical potentials. The corresponding 
study, when extrapolated to the regime of realistic densities, may further extend our understanding of 
dense quark matter by clarifying (i) possible directional dependences of the gap function, (ii) the evolution 
of such a dependence between the two limiting cases studied here, and (iii) the effect of $\beta$-equilibrium 
and neutrality of quark matter on the gap function in magnetic fields. All of these topics are left for future 
investigations.

\section*{Acknowledgments}

The authors would like to thank E.~Gorbar for interesting discussions and useful comments. 
This work is supported in part by the U.S. National Science Foundation under Grant No. PHY-0969844.

\appendix
\section{Quark propagator}
\label{Appendix A}

In this appendix we calculate the explicit forms of the full propagators for quasiparticles with 
$\tilde{Q}=+\frac{1}{2}$ charge. (The result can be also easily generalized to quasiparticles 
with $\tilde{Q}=-\frac{1}{2}$ charge.) We present the details of the analysis for $11$- and 
$21$-components of the propagator $S_{(+\frac{1}{2})}^X$.

The starting point of the derivation is the definition of the inverse propagator in Eq.~(\ref{diagS1/2^1}). 
Introducing a shorthand notation for the diagonal and off-diagonal elements of that propagator, we 
write
\begin{equation}
S_{(+\frac{1}{2})}^X = \left(
         \begin{array}{cc}
                      [G^+_0]^{-1} & \Delta^- \\
                      \Delta^+ & [G^-_0]^{-1} \\
                    \end{array}
                  \right)^{-1} =
\left(
  \begin{array}{cc}
    G^+ & \Xi^- \\
    \Xi^+ & G^- \\
  \end{array}
\right),
\end{equation}
where  
\begin{eqnarray}
  G^{\pm} &=& [(G^{\pm}_0)^{-1}-\Delta^{\mp}G^{\mp}_0\Delta^{\pm}]^{-1} ,\\
  \Xi^{\pm} &=& -G^{\mp}_0\Delta^{\pm}G^{\pm}.
\end{eqnarray}
The explicit forms of the $11$- and $21$-components of the propagator read
\begin{eqnarray}\label{S(1/2)^1_11}
  S_{(+\frac{1}{2})11}^{X}
   &=&\left( \gamma^{\mu} \pi^{(+\frac{1}{2})}_{\mu}-\mu_{dg}\gamma^0+m\right)
   \bigg[\left( \gamma^{\mu} \pi^{(+\frac{1}{2})}_{\mu} +\mu_{ur}\gamma^0-m\right)\nonumber\\
   &&\times 
   \left( \gamma^{\mu} \pi^{(+\frac{1}{2})}_{\mu}-\mu_{dg}\gamma^0+m\right)-\Delta^2\bigg]^{-1},\\
\label{S(1/2)^1_21}
  S_{(+\frac{1}{2})21}^{X}
  &=& -i\gamma^5\Delta^* \left[\left( \gamma^{\mu} \pi^{(+\frac{1}{2})}_{\mu}+\mu_{ur}\gamma^0-m\right)
  \left( \gamma^{\mu} \pi^{(+\frac{1}{2})}_{\mu}-\mu_{dg}\gamma^0+m\right)-\Delta^2\right]^{-1},
\end{eqnarray}
where, by definition, $\pi^{(\tilde{Q})}_{\mu} \equiv  i\partial_{\mu} + \tilde{e}\tilde{Q}\tilde{A}_{\mu}$ and 
the gauge field is $\tilde{A}^{\mu}=(0,0,x\tilde{B},0)$ with the strength of the external (rotated) magnetic 
field denoted by $\tilde{B}$.

The inverse of the operator in the square brackets of Eqs.~(\ref{S(1/2)^1_11}) and (\ref{S(1/2)^1_21}), 
which is the same for all components of the propagator, can be calculated by employing the usual trick of 
``quadrating" the operator. In this case, however, we end up ``biquadrating" it because the corresponding 
operator is already quadratic in energy. For this purpose, let us introduce the following shorthand notation:
\begin{eqnarray}\label{X1}
  \hat{X}^{\pm} &=& \left[(i\partial_t- \delta\mu )^2-\bm{\pi}_{\perp}^2-i\tilde{e}\tilde{Q}\tilde{B}\gamma^1\gamma^2
  -(\pi^3)^2-m^2-\bar{\mu}^2 -\Delta^2 \right] \nonumber\\
  &\pm&  2 \gamma^0\bar{\mu} \left(\bm{\gamma}_{\perp}\cdot\bm{\pi}_{\perp}+\gamma^3\pi^3 -m\right),
\end{eqnarray}
where $\delta\mu = \frac{\mu_{dg}-\mu_{ur}}{2}$, $\bar{\mu} = \frac{\mu_{ur}+\mu_{dg}}{2}$,
$\bm{\pi}_{\perp}=(\pi^1,\pi^2)$ and $\bm{\gamma}_{\perp}=(\gamma^1, \gamma^2)$. Note that $\hat{X}^{-}$ 
is the same operator that appears in the square brackets of Eqs.~(\ref{S(1/2)^1_11}) and (\ref{S(1/2)^1_21}). 
For simplicity of notation, we dropped index $\tilde{Q}$ here.

Let us first concentrate on the $11$-component of the propagator. It can be rewritten as follows:
\begin{eqnarray}\label{SX_11}
  S_{(+\frac{1}{2})11}^{X}
   &=&\left( \gamma^{\mu} \pi^{(+\frac{1}{2})}_{\mu}-\mu_{dg}\gamma^0+m\right)\hat{X}^{+}
         \left( \hat{X}^{-} \hat{X}^{+}\right)^{-1}   \nonumber\\
   &\equiv & \left(\hat{A}-\bm{\gamma}_{\perp}\cdot\bm{\pi}_{\perp}\hat{B}\right) \hat{C}^{-1} .
\end{eqnarray}
The three new operator functions introduced here are defined by 
\begin{eqnarray}
\hspace*{-7mm}
\hat{A}&=& \!\!  \left[(i\partial_t)\gamma^0-\pi^3\gamma^3-\mu_{dg}\gamma^0+m\right]
\left[\left(i\partial_t- \delta\mu \right)^2-\bar{\mu}^2 -2\bar{\mu} (\gamma^3\pi^3+m)\gamma^0
-(\pi^3)^2  -m^2 -\Delta^2\right]  \nonumber \\
   &-& \left[(i\partial_t)\gamma^0-\pi^3\gamma^3+\mu_{ur}\gamma^0+m\right] 
  \left(\bm{\pi}_{\perp}^2+i\tilde{e}\tilde{Q}\tilde{B}\gamma^1\gamma^2\right), \label{A}\\
\hat{B}&=&  (i\partial_t-\mu_{dg})^2  -\bm{\pi}_{\perp}^2-i\tilde{e}\tilde{Q}\tilde{B}\gamma^1\gamma^2
-(\pi^3)^2-m^2  -\Delta^2,  \label{B}\\
\hat{C}&=& \left[\left(i\partial_t- \delta\mu \right)^2-\bm{\pi}_{\perp}^2-i\tilde{e}\tilde{Q}\tilde{B}\gamma^1\gamma^2
    -(\pi^3)^2-m^2+\bar{\mu}^2-\Delta^2\right]^2 \nonumber \\
   &-& 4 \bar{\mu}^2 \left[\left(i\partial_t- \delta\mu \right)^2-\Delta^2\right].  \label{C}
\end{eqnarray}
In the coordinate space, the corresponding propagator is formally given by
\begin{equation}
\label{S(1/2)11^1}
    S_{(+\frac{1}{2})11}^{X}(u,u^\prime)=\langle u|\left(\hat{A}-\bm{\gamma}_{\perp}\cdot\bm{\pi}_{\perp}\hat{B}\right)
   \hat{C}^{-1} |u' \rangle,
\end{equation}
where $u=(t,z,\bm{r}_{\perp})$ and $\bm{r}_{\perp}=(x,y)$. It is easy to perform a Fourier transform 
in time and $z$-coordinate,
\begin{equation}\label{S(1/2)11^1(r)}
  S_{(+\frac{1}{2})11}^{X}(\omega,k^3;\bm{r}_{\perp},\bm{r}'_{\perp}) =
  \int dt \, dz \, e^{i\omega t-ik^3z} \, S_{(+\frac{1}{2})11}^1(u,u^\prime) .
\end{equation}
In essence, this transform results in a simple replacement of $i\partial_t \to \omega$ and $\pi^3 \to k^3$ 
in all of the earlier expressions.

To proceed further, we should find a basis of suitable eigenstates, in which the propagator has the simplest 
possible form. To this end, we note that the functions $\hat{A}$, $\hat{B}$ and $\hat{C}$ depend on the 
operator $\bm{\pi}_{\perp}^2+i\tilde{e}\tilde{Q}\tilde{B}\gamma^1\gamma^2$. Its eigenvalues 
are well known: $2n|\tilde{e}\tilde{Q}\tilde{B}|$, where $n=0,1,2,\ldots$ is the Landau-level index. 
Note that the integer quantum number $n$ has both the orbital and spin contributions, i.e., $n=k+(1+s)/2$, 
where $k=0,1,2,\ldots$ labels a specific orbital state, while $s=\pm 1$ corresponds to a given (up or down) 
spin state. The explicit form of the corresponding eigenstates $\langle \bm{r}_{\perp}| k\, p_y\, s \rangle$ is 
also well known (e.g., see Ref.~\cite{Gorbar2011}, where similar method and notations are used).

Following closely the approach of Ref.~\cite{Gorbar2011}, we use the complete set of eigenstates to 
simplify the expression for the propagator (\ref{S(1/2)11^1(r)}). The final result will have the form 
\begin{equation} 
  S_{(+\frac{1}{2})21}^{X}(\omega,k^3;\bm{r}_{\perp},\bm{r}'_{\perp})
  = e^{i\Phi(\bm{r}_{\perp},\bm{r}'_{\perp})}\bar{S}_{(+\frac{1}{2})21}^{X}(\omega,k^3;\bm{r}_{\perp}-\bm{r}'_{\perp}),
  \label{S+1/2,21}
\end{equation}
where $\Phi(\bm{r}_{\perp},\bm{r}'_{\perp})$ is the Schwinger phase.  In the Landau gauge used, the explicit form 
of the phase is 
\begin{equation}\label{Phi}
    \Phi(\bm{r}_{\perp},\bm{r}'_{\perp})=-\frac{(x+x')(y-y')}{2l^2}\mbox{sgn}(\tilde{e}\tilde{Q}\tilde{B}),
\end{equation}
where $l=1/\sqrt{|\tilde{e}\tilde{Q}\tilde{B}|}$ is the magnetic length. (Note that this phase is responsible for 
breaking the translational invariance of the propagator.) The translationally invariant part of the  
propagator is given by 
\begin{eqnarray}\label{barS(1/2)11^1(r)}
\hspace*{-7mm}
  \bar{S}_{(+\frac{1}{2})11}^{X}(\omega,k^3;\bm{r}_{\perp} ) &=& \frac {e^{-\xi/2}}{2\pi l^2}
  \sum_{n=0}^{\infty}\left\{\frac{\mathcal{A}_n}{\mathcal{C} _n  } \left[L_n(\xi)\mathcal{P}_{-}+L_{n-1}(\xi)\mathcal{P}_{+}\right]
  -i\frac{\bm{\gamma}_{\perp}\cdot \bm{r}_{\perp}}{l^2}
   \frac{\mathcal{B}_n} {\mathcal{C} _n }L^1_{n-1}(\xi)\right\},
\end{eqnarray}
where $\xi\equiv \bm{r}_{\perp}^2/(2l^2)$, $L_n^{\alpha}(\xi)$ are the generalized
Laguerre polynomials (by definition, $L_n\equiv L_n^0$ and $L^{\alpha}_{-1}=0$),
and
\begin{equation}\label{P}
    \mathcal{P}_{\pm}=\frac{1}{2}\left(1\pm i\gamma^1\gamma^2 \mbox{sgn}(\tilde{e}\tilde{Q}\tilde{B})\right)
\end{equation}
are the spin projection operators. 

Functions $\mathcal{A}_n$, $\mathcal{B}_n$ and $\mathcal{C} _n$ in Eq.~(\ref{barS(1/2)11^1(r)}) 
replace the corresponding operators $\hat{A}$, $\hat{B}$ and $\hat{C}$, when projected onto the 
$n$th Landau-level state. Their explicit forms are obtained from $\hat{A}$, $\hat{B}$ and $\hat{C}$ 
by replacing $\bm{\pi}_{\perp}^2+i\tilde{e}\tilde{Q}\tilde{B}\gamma^1\gamma^2\to 2n|\tilde{e}\tilde{Q}\tilde{B}|$, 
i.e.,
\begin{eqnarray}
        \mathcal{A}_n
  &=&\left[\omega\gamma^0-k^3\gamma^3-\mu_{dg}\gamma^0+m\right]
\left[\left(\omega- \delta\mu \right)^2-\bar{\mu}^2 -2\bar{\mu} (\gamma^3k^3+m)\gamma^0
-(k^3)^2  -m^2 -\Delta_n^2\right]  \nonumber \\
   &-& 2n|\tilde{e}\tilde{Q}\tilde{B}|\left[\omega\gamma^0-k^3\gamma^3+\mu_{ur}\gamma^0+m\right] ,   
   \label{A-n} \\
        \mathcal{B}_n
  &=&  (\omega-\mu_{dg})^2  -2n|\tilde{e}\tilde{Q}\tilde{B}|
-(k^3)^2-m^2  -\Delta_n^2,
   \label{B-n}  \\
        \mathcal{C}_n
  &=&  \left[\left(\omega- \delta\mu \right)^2-2n|\tilde{e}\tilde{Q}\tilde{B}|
    -(k^3)^2-m^2+\bar{\mu}^2-\Delta_n^2\right]^2 
   - 4 \bar{\mu}^2 \left[\left(\omega- \delta\mu \right)^2-\Delta_n^2\right].
   \label{C-n}
\end{eqnarray}
Here we consider a general case when the dynamically generated gap function $\Delta_n$ depends 
not only on the energy $\omega$ and $k^3$, but also on the Landau-level index $n$. (In operator form, 
it means that $\Delta$ depends on $\bm{\pi}_{\perp}^2+i\tilde{e}\tilde{Q}\tilde{B}\gamma^1\gamma^2$.) 
Therefore, we replaced the operator $\Delta$ with the corresponding value $\Delta_n$ that it takes in the 
$n$th Landau-level state. 

At this point, it may be appropriate to note that the zeros of $\mathcal{C}_n$ determine the
spectrum of quasiparticles in color superconducting quark matter in a magnetic field, i.e.,
\begin{equation} 
E_{n,\pm,\pm}  =  \delta\mu  \pm \sqrt{\left[\sqrt{2n |\tilde{e}\tilde{Q}\tilde{B}| + (k^3)^2+m^2}  
\pm \bar{\mu}\right]^2+\Delta_n^2}.
\end{equation}
Note that all four different sign combinations are possible. The choice of the sign in front of the chemical 
potential $\bar{\mu}$ corresponds to the choice of either particle states (allowing small energies of order 
$\Delta_n$) or antiparticle states (generally having large energies of order $\bar{\mu}$). The sign in 
front of the overall square root corresponds to particle/hole type quasiparticles (i.e., positive/negative 
energy states). One should note, however, that an additional complication in this classification 
appears in the case of gapless superconducting phases when $ \delta\mu  > \Delta_n$ \cite{g2SC,gCFL}.

Following the same approach, we can derive explicit expressions for all components of the 
propagator $S_{(+\frac{1}{2})}^{X}$. For example, the final expression for the off-diagonal 
$21$-component, which is used in the gap equation in the main text, reads
\begin{equation}\label{S(1/2)21^1(r)1}
  S_{(+\frac{1}{2})21}^{X}(\omega,k^3;\bm{r}_{\perp},\bm{r}'_{\perp})
  = e^{i\Phi(\bm{r}_{\perp},\bm{r}'_{\perp})}\bar{S}_{(+\frac{1}{2})21}^{X}(\omega,k^3;\bm{r}_{\perp}-\bm{r}'_{\perp}),
\end{equation}
with the translationally invariant part given by 
\begin{equation}\label{barS(1/2)21^1(r)}
  \bar{S}_{(+\frac{1}{2})21}^{X}(\omega,k^3;\bm{r}_{\perp} )= -i\gamma^5\frac {e^{-\xi/2}}{2\pi l^2}
  \sum_{n=0}^{\infty} \Delta^*_n \bigg\{\frac{\mathcal{E}_n}
  {\mathcal{C} _n} [L_n(\xi)\mathcal{P}_{-}+L_{n-1}(\xi)\mathcal{P}_{+}]   -i\frac{\bm{\gamma}_{\perp}\cdot \bm{r}_{\perp}}{l^2}
   \frac{2\bar{\mu} \gamma^0}
  {\mathcal{C} _n}L^1_{n-1}(\xi) \bigg\}.
\end{equation}
Here we introduced yet another function, 
\begin{equation}
\mathcal{E}_n
  = (\omega- \delta\mu )^2-2n|\tilde{e}\tilde{Q}\tilde{B}|-(k^3)^2-m^2-\bar{\mu}^2 
  -\Delta_n^2 -2\bar{\mu} (k^3\gamma^3+m)\gamma^0.
   \label{E-n}
\end{equation}
Before concluding this appendix, let us add that similar representations can be also derived for 
the components of the inverse propagator. As an example, let us present the corresponding result for 
$\left[S^X_{(+\frac{1}{2})}\right]^{-1}_{21}(u,u^\prime)$, which is used in the gap equation. It has the same 
general structure as the above expressions for the components of $S_{(+\frac{1}{2})}^{X}$, i.e., 
\begin{equation}
\left[S^X_{(+\frac{1}{2})}\right]^{-1}_{21}(\omega,k^3;\bm{r}_{\perp} ,\bm{r}'_{\perp})
= e^{i\Phi(\bm{r}_{\perp},\bm{r}'_{\perp})}\left[\overline{S^{X}}_{(+\frac{1}{2})}\right]_{21}^{-1}(\omega,k^3;\bm{r}_{\perp}-\bm{r}'_{\perp}).
\end{equation}
It is important that the inverse propagator has exactly the same phase as the propagator itself;
see Eqs.~(\ref{S+1/2,21}) and (\ref{Phi}). The explicit form of its translationally invariant part reads
\begin{equation}
\left[\overline{S^{X}}_{(+\frac{1}{2})}\right]_{21}^{-1}(\omega,k^3;\bm{r}_{\perp} )
= -i\gamma^5 \frac{e^{-\xi/2}}{2\pi l^2}
  \sum_{n=0}^{\infty}\Delta^*_n\left[L_n(\xi)\mathcal{P}_{-}+L_{n-1}(\xi)\mathcal{P}_{+}\right].
  \label{barDelta+}
\end{equation}

\section{Gap equation}
\label{Appendix B}

The gap equation (i.e., the off-diagonal component of the Schwinger-Dyson equation 
for the full propagator) in the coordinate space reads
\begin{equation} 
\left[S^X_{(\tilde{Q})}\right]^{-1}_{21}(u,u^\prime) =-ig^2\gamma^{\mu} \left(T^A\right)^T \left[S^X_{(\tilde{Q})}\right]_{21}(u,u^\prime) 
\gamma^{\nu}T^B D^{AB}_{\mu\nu}(u-u') ,
\end{equation}
where $D^{AB}_{\mu\nu}(u,u^\prime)$ is the gluon propagator, which is assumed to be diagonal in 
adjoint color indices ($A, B=1,2,...,8$), i.e., $D^{AB}_{\mu\nu}(u-u')=\delta^{AB}D_{\mu\nu}(u-u')$. 
By making use of the identity
\begin{eqnarray}
  \sum^8_{A=1} T^A_{a'a} T^A_{b'b} &=& \frac{1}{2}\delta_{a'b}\delta_{ab'}-\frac{1}{6}\delta_{aa'}\delta_{bb'},
\end{eqnarray}
we derive the following form of the gap equation:
\begin{equation} 
     \left[S^X_{(\tilde{Q})}\right]^{-1}_{21}(u,u^\prime)=i\frac{2}{3}g^2\gamma^{\mu} 
     \left[S^X_{(\tilde{Q})}\right]_{21}(u,u^\prime)\gamma^{\nu} D_{\mu\nu}(u-u').
\end{equation}
Taking into account that all components of the quasiparticle propagator as well as its inverse have 
the same nonzero Schwinger phase, we can derive the equation for the translationally invariant
parts simply by dropping the common phase factor on both side of the gap equation,
\begin{eqnarray}\label{gap+1/2 transverse}
\left[\overline{S^{X}}_{(+\frac{1}{2})}\right]_{21}^{-1}(\omega,k^3;\bm{r}_{\perp} )
  &=& i\frac{2g^2}{3}\int\frac{d\omega' dk'^3}{(2\pi)^2} \gamma^{\mu}
  \bar{S}_{(+\frac{1}{2})21}^{X}(\omega,k^3;\bm{r}_{\perp})\gamma^{\nu} \nonumber\\
  &&\times\int \frac{d^2\bm{q}_{\perp}}{(2\pi)^2}e^{i\bm{q}_{\perp}\cdot\bm{r}_{\perp}}
  D_{\mu\nu}(\omega-\omega',k^3-k'^3,\bm{q}_{\perp}),
\end{eqnarray}
where we additionally performed a Fourier transform in time and $z$-coordinate on both sides 
of the equation, and used a momentum representation for the gluon propagator. 

By making use of the explicit form of the relevant translationally invariant parts of the propagators in 
Eqs.~(\ref{barS(1/2)21^1(r)}) and (\ref{barDelta+}), we rewrite the last form of the gap equation as follows:
\begin{eqnarray}
 \frac{e^{-\xi/2}}{2\pi l^2}
  \sum_{n=0}^{\infty}\Delta_n[L_n(\xi)\mathcal{P}_{-} &+& L_{n-1}(\xi)\mathcal{P}_{+}]  
  = -i\frac{2g^2}{3} \frac {e^{-\xi/2}}{2\pi l^2} \sum_{n=0}^{\infty} \int\frac{d\omega' dk'^3}{(2\pi)^2}
  \gamma^{\mu}\frac{\Delta_n}{\mathcal{C} _n}  \nonumber\\
&\times  &   \left\{
  \mathcal{E}_n\left[L_n(\xi)\mathcal{P}_{-}+L_{n-1}(\xi)\mathcal{P}_{+}\right] - 2\bar{\mu} i\frac{\bm{\gamma}_{\perp}\cdot \bm{r}_{\perp}}{l^2} 
   \gamma^0 L^1_{n-1}(\xi)
  \right\}\gamma^{\nu}\nonumber\\
&\times &   \int \frac{d^2\bm{q}_{\perp}}{(2\pi)^2} e^{i\bm{q}_{\perp}\cdot\bm{r}_{\perp}}
D_{\mu\nu}(\omega-\omega',k^3-k'^3,\bm{q}_{\perp}).
\end{eqnarray}
The last equation can now be easily projected onto different orbital eigenstates. This is formally done by multiplying 
both sides of the equation by $e^{-\xi/2} L_m(\xi)$ (where $m=0,1,2,\ldots$) and integrating over the perpendicular 
spatial coordinates $\bm{r}_{\perp}$. After performing such projections, we arrive at the following (infinite) set of 
gap equations in the Landau-level representation:
\begin{eqnarray}\label{Full gap A}
 \Delta_m\mathcal{P}_{-}+\Delta_{m+1}\mathcal{P}_{+} 
  &=& -i\frac{2g^2}{3}\sum_{n=0}^{\infty}\int\frac{d\omega' dk'^3}{(2\pi)^2}
   \int \frac{d^2\bm{q}_{\perp}}{(2\pi)^2}\gamma^{\mu} \frac{\Delta_n\mathcal{E}_n}{\mathcal{C} _n}
   \bigg[ \mathcal{L}^{(0)}_{n,m} \left(\frac{q_{\perp}^2l^2}{2}\right) \mathcal{P}_{-} \nonumber\\
  &&     +\mathcal{L}^{(0)}_{n-1,m} \left(\frac{q_{\perp}^2l^2}{2}\right) \mathcal{P}_{+} \bigg]\gamma^{\nu}
  D_{\mu\nu}(\omega-\omega',k^3-k'^3,\bm{q}_{\perp}),
\end{eqnarray}
where, by definition, 
\begin{equation}\label{L(0)}
    \mathcal{L}^{(0)}_{n,m}\left(x\right)=(-1)^{n+m}e^{-x}
    L^{m-n}_n\left(x\right)L^{n-m}_m\left(x\right).
\end{equation}
In the derivation, we used the following table integrals (see formulas 7.414 3 and 7.422 2  in Ref.~\cite{Gradshteyn2000}):
\begin{eqnarray}
\int^{\infty}_{0} dx e^{-x}x^{\alpha}L^{\alpha}_{m}(x)L^{\alpha}_{n}(x)&=& \frac{\Gamma(n+\alpha+1)}{n!} \delta^{n}_{m},
\end{eqnarray}
and
\begin{equation}
  \int^{\infty}_{0} \!\!\!\! dx x^{2\sigma+1}e^{-\alpha x^2}
  L^{\sigma}_{m}(\alpha x^2)L^{\sigma}_{n}(\alpha x^2) J_0(xy) =
  \frac{(-1)^{m+n}}{2\alpha^{\sigma+1}} \frac{(m+\sigma)!}{m!}e^{-y^2/4\alpha}
  L^{n-m}_{m+\sigma}\bigg(\frac{y^2}{4\alpha}\bigg)L^{m-n}_n\bigg(\frac{y^2}{4\alpha}\bigg). 
\end{equation}

\section{Propagator in weak magnetic field limit}
\label{Appendix C}

In this Appendix, we consider the quasiparticle propagator and the gap equation in the limit 
of weak magnetic field. 

We begin by performing a Fourier transform of the translation invariant part of the propagator,
\begin{eqnarray}\label{barS(1/2)21^1(k)}
  \bar{S}_{(+\frac{1}{2})21}^{X}(\omega,k^3,\bm{k}_{\perp})
  &=&\int d^2\bm{r}_{\perp} e^{-i\bm{k}_{\perp} \cdot \bm{r}_{\perp}}
  \bar{S}_{(+\frac{1}{2})21}^{X}(\omega,k^3;\bm{r}_{\perp})\nonumber \\
  &=&-2i\gamma^5 e^{-k_\perp^2l^2}
  \sum_{n=0}^{\infty}(-1)^n
  \frac{\Delta_n}{\mathcal{C} _n}
  \bigg\{ \mathcal{E}_n[L_n\left(2k_\perp^2l^2\right)\mathcal{P}_{-}-L_{n-1}\left(2k_\perp^2l^2\right)\mathcal{P}_{+}] \nonumber \\
   &&+4\bar{\mu}(\bm{\gamma}_{\perp}\cdot \bm{k}_{\perp})\gamma^0 L^1_{n-1}\left(2k_\perp^2l^2\right) \bigg\}.
\end{eqnarray}
In the weak field limit, the difference between the neighboring levels is vanishingly small in energy 
and the properties of the corresponding states become almost indistinguishable. In application to the 
gap function $\Delta_n$, this means that it will become almost independent of the Landau-level index 
in a wide range of $n$ near the (would be) Fermi surface. (Strictly speaking, the true Fermi surface is 
not well defined in a superconductor, but if the gap is small, $\Delta\ll\bar{\mu}$, one could map the 
corresponding phase space onto the phase space in the free quark matter.)

In order to derive a weak field expression for the propagator, one needs to first perform the sum 
over the Landau-level index $n$. A straightforward way of achieving this is to employ the usual 
proper-time representation, i.e.,
\begin{eqnarray}
\frac{1}{(a+2n|b|)^2+c^2} &=&  \int_{0}^{\infty}\frac{ds}{c} \sin(s c) e^{-s(a+2n|b|)},
\label{proper-time-1} \\
\frac{a+2n|b|}{(a+2n|b|)^2+c^2} &=&  \int_{0}^{\infty} ds  \cos(s c) e^{-s(a+2n|b|)},
\label{proper-time-2} 
\end{eqnarray}
for the two types of structures appearing in the Euclidian propagator, and then use the well 
known summation formula for Laguerre polynomials,
\begin{equation}
 \sum_{n=0}^{\infty}L^{\alpha}_n(x)z^n= (1-z)^{-(\alpha+1)}\exp\left(\frac{xz}{z-1}\right).
 \label{LaguerreSum}
\end{equation}
Before using these identities, it is convenient to rewrite propagator (\ref{barS(1/2)21^1(k)}) in 
the following form:
\begin{equation}\label{barS(1/2)Euclidian}
  \bar{S}_{(+\frac{1}{2})21}^{X}(i\omega_E,k^3,\bm{k}_{\perp})
  =i\gamma^5 \Delta \left[ I_{1}
   + 2 \bar{\mu} (k^3\gamma^3+m+\bar{\mu}\gamma^0)\gamma^0 I_{2}
   + 2 \bar{\mu} (\bm{\gamma}_{\perp}\cdot \bm{k}_{\perp})\gamma^0 I_{3}\right],
\end{equation}
where, by definition, the sums $I_{i}$ ($i=1,2,3$) are  
\begin{eqnarray} 
I_{1}  &=& 2 e^{-\frac{k_\perp^2}{|b|}} \sum_{n=0}^{\infty}(-1)^n  L_n\left(\frac{2k_\perp^2}{|b|}\right)
\left(\frac{a+2n|b|}{(a+2n|b|)^2+c^2}\mathcal{P}_{-} 
 +\frac{a+2(n+1)|b|}{[a+2(n+1)|b|]^2+c^2}\mathcal{P}_{+} \right) ,
\\
I_{2}  &=& 2  e^{-\frac{k_\perp^2}{|b|}} \sum_{n=0}^{\infty}(-1)^n  L_n\left(\frac{2k_\perp^2}{|b|}\right)
\left(\frac{1}{(a+2n|b|)^2+c^2}\mathcal{P}_{-} 
 +\frac{1}{[a+2(n+1)|b|]^2+c^2}\mathcal{P}_{+} \right) ,
\\
I_{3}  &=& 4  e^{-\frac{k_\perp^2}{|b|}} \sum_{n=0}^{\infty}(-1)^n  L^1_n\left(\frac{2k_\perp^2}{|b|}\right)
\frac{1}{[a+2(n+1)|b|]^2+c^2}.
\end{eqnarray}
Here we used the following notation:
\begin{eqnarray}\label{someNotation}
  a &=& (\omega_E+i \delta\mu )^2 +(k^3)^2 +m^2+\Delta^2-\bar{\mu}^2 ,\\
  b &=& \tilde{e}\tilde{Q}\tilde{B} ,\\
  c &=& 2\bar{\mu}\sqrt{ (\omega_E+i \delta\mu )^2+\Delta^2}.
\end{eqnarray}
It is appropriate to mention that the use of the proper-time representations, as given 
by Eqs.~(\ref{proper-time-1}) and (\ref{proper-time-2}), may not be completely justified 
in the presence of a nonzero density. Indeed, when the chemical potential is sufficiently 
large, the above expression for the parameter $a$ may become negative. When this occurs, 
the proper-time integrals become divergent and the validity of the derivation seemingly 
fails. The way around this problem is to assume that the chemical potential is sufficiently 
small at all intermediate stages of derivation. In the end, after magnetic field expansion 
is done and all proper-time integrations are performed, one can extend the validity of the 
propagators to large values of the chemical potential. 

With the above remark kept in mind, we use the proper-time representations to rewrite 
the expressions for the sums $I_{i}$ as follows:
\begin{eqnarray}
I_{1} &=& 2 e^{-k_\perp^2/|b|} \sum_{n=0}^{\infty}(-1)^n  L_n\left(\frac{2k_\perp^2}{|b|}\right) 
   \int_{0}^{\infty} ds  \cos(s c) e^{-s(a+2n|b|)}
   \left( \mathcal{P}_{-} +e^{-2 |b| s}\mathcal{P}_{+}  \right), \\
I_{2} &=& 2 e^{-k_\perp^2/|b|} \sum_{n=0}^{\infty}(-1)^n  L_n\left(\frac{2k_\perp^2}{|b|}\right) 
   \int_{0}^{\infty} \frac{ds}{c}  \sin(s c) e^{-s(a+2n|b|)}
   \left( \mathcal{P}_{-} +e^{-2 |b| s}\mathcal{P}_{+}  \right) ,\\
I_{3} &=& 4 e^{-k_\perp^2/|b|} \sum_{n=0}^{\infty}(-1)^n  L^1_n\left(\frac{2k_\perp^2}{|b|}\right) 
   \int_{0}^{\infty} \frac{ds}{c}  \sin(s c) e^{-s(a+2n|b|+2|b|)}.
\end{eqnarray}
Then, after using the summation formula (\ref{LaguerreSum}), we derive
\begin{eqnarray}
I_{1} &=& \int_{0}^{\infty} ds  \cos(s c) e^{-sa-(k_\perp^2/b) \tanh(sb)}\left[1-i\gamma^1\gamma^2 \tanh(sb)\right], \\
I_{2} &=& \int_{0}^{\infty} \frac{ds}{c}  \sin(s c) e^{-sa-(k_\perp^2/b) \tanh(sb)}\left[1-i\gamma^1\gamma^2 \tanh(sb)\right],\\
I_{3} &=&\int_{0}^{\infty} \frac{ds}{c}  \sin(s c) e^{-sa-(k_\perp^2/b) \tanh(sb)}
\frac{1}{\cosh^2(sb)} .
\end{eqnarray}
Finally, expanding the integrands in powers of the magnetic field $b$ and integrating over the proper time, 
we obtain
\begin{eqnarray}
I_{1} &\simeq & \int_{0}^{\infty} ds  \cos(s c) e^{-s(a+k_\perp^2)}
      \left(1- i \gamma^1\gamma^2 s b +\frac{s^3}{3} k_\perp^2 b^2 +O(b^3)\right)
      = \frac{a+k_\perp^2}{(a+k_\perp^2)^2+c^2}       \nonumber \\
   &-&  i \gamma^1\gamma^2\frac{(a+k_\perp^2)^2-c^2}{[(a+k_\perp^2)^2+c^2]^2}b
+\frac{2[(a+k_\perp^2)^4-6(a+k_\perp^2)^2c^2+c^4] k_\perp^2 }
{[(a+k_\perp^2)^2+c^2]^4}b^2 +O(b^3), \\
I_{2} &\simeq & \int_{0}^{\infty} \frac{ds}{c}  \sin(s c) e^{-s(a+k_\perp^2)}
      \left(1- i \gamma^1\gamma^2 s b +\frac{s^3}{3} k_\perp^2 b^2 +O(b^3)\right)
      =  \frac{1}{(a+k_\perp^2)^2+c^2} \nonumber \\
   &-& i \gamma^1\gamma^2\frac{2(a+k_\perp^2)}{[(a+k_\perp^2)^2+c^2]^2}b
+\frac{8 (a+k_\perp^2) [(a+k_\perp^2)^2-c^2] k_\perp^2 }
{[(a+k_\perp^2)^2+c^2]^4}b^2 +O(b^3),\\
I_{3} &\simeq & \int_{0}^{\infty} \frac{ds}{c}  \sin(s c) e^{-s(a+k_\perp^2)}
      \left(1 -s^2 b^2 + \frac{s^3}{3} k_\perp^2 b^2 +O(b^3)\right)=  \frac{1}{(a+k_\perp^2)^2+c^2} \nonumber \\
   &-& \frac{2[3(a+k_\perp^2)^2-c^2] }{[(a+k_\perp^2)^2+c^2]^3}b^2 
+\frac{8 (a+k_\perp^2) [(a+k_\perp^2)^2-c^2] k_\perp^2 }
{[(a+k_\perp^2)^2+c^2]^4}b^2 +O(b^3).
\end{eqnarray}
Now, combining the same order terms in powers of the magnetic field, we rewrite propagator 
(\ref{barS(1/2)Euclidian}) as follows:
\begin{equation}\label{barS(1/2)-Ks}
  \bar{S}_{(+\frac{1}{2})21}^{X}(i\omega_E,k^3,\bm{k}_{\perp})
  =i\gamma^5 \Delta \left[ K^{(0)}  +K^{(1)} +K^{(2)} \right],
\end{equation}
where 
\begin{eqnarray}
K^{(0)} &=&   \frac{a^{+}_k+a^{-}_k+4 \bar{\mu} (\bm{\gamma}\cdot \bm{k}+m)\gamma^0}{2a_k^{+}a^{-}_k} ,
\label{K^(0)}\\
K^{(1)} &=& - i \gamma^1\gamma^2
\frac{(a^{+}_k)^2+(a^{-}_k)^2-4\bar{\mu}^2 (a^{+}_k+a^{-}_k)+8\bar{\mu} a_k (k^3\gamma^3+m)\gamma^0}
{2 (a_k^{+}a^{-}_k)^2}b ,
\label{K^(1)}\\
K^{(2)} &=&  \frac{a^{+}_k(a^{+}_k-4\bar{\mu}^2)^3+a^{-}_k(a^{-}_k-4\bar{\mu}^2)^3
+4\bar{\mu}^2a^{+}_ka^{-}_k\left[16\bar{\mu}^2-3(a^{+}_k+a^{-}_k)\right]}{(a_k^{+}a^{-}_k)^4} k_\perp^2  b^2 
\nonumber\\
&-& 4 \bar{\mu} (\bm{\gamma}_{\perp}\cdot \bm{k}_{\perp})\gamma^0
\frac{ 4a_k^2-a_k^{+}a^{-}_k}{(a_k^{+}a^{-}_k)^3}b^2 + 16 \bar{\mu} a_k (\bm{\gamma}\cdot \bm{k}+m)\gamma^0 
\frac{ 2a_k^2-a_k^{+}a^{-}_k} {(a_k^{+}a^{-}_k)^4} k_\perp^2  b^2.
\label{K^(2)}
\end{eqnarray}
Note the shorthand notation used,
\begin{eqnarray}
\bm{\gamma}\cdot \bm{k} &\equiv & \bm{\gamma}_{\perp}\cdot \bm{k}_{\perp}+k^3\gamma^3,\\
a_k^{\pm} &\equiv &  (\omega_E+i \delta\mu )^2 +\left(E_{k}\pm\bar{\mu}\right)^2 +\Delta^2,\\
a_k &\equiv &  a+k_\perp^2 =  (\omega_E+i \delta\mu )^2 +k^2 +m^2+\Delta^2-\bar{\mu}^2,
\end{eqnarray}
as well as $E_{k}\equiv  \sqrt{k^2 +m^2}$ and $k^2 \equiv  k_\perp^2 +(k^3)^2$.

\section{Gap equation in weak magnetic field limit}
\label{Appendix D}

To leading order (i.e., the limit of vanishing magnetic field), the gap equation reads
\begin{equation}
\Delta^{(0)}(\omega_E)=\frac{g^2}{6}\int\frac{d\omega_E^\prime}{2\pi}
   \int \frac{d^3\bm{k}^\prime}{(2\pi)^3}   \Delta^{(0)}(\omega_E^\prime) 
   \mbox{tr} \left[ \gamma^{\mu}K^{(0)}(\omega^\prime,\bm{k}^\prime) \gamma^{\nu}\right] 
   D_{\mu\nu}(\omega-\omega^\prime,\bm{k}-\bm{k}^\prime). 
\end{equation}
Here we assumed that the gap is an explicit function of the energy, but not of the momentum. 
The result for the trace in the integrand is given by
\begin{equation}
\mbox{tr}\left[ \gamma^{\mu}K^{(0)}(\omega^\prime,\bm{k}^\prime)\gamma^{\nu}\right] 
= 2 g^{\mu\nu}  \frac{a^{+}_{k^\prime}+a^{-}_{k^\prime}}{a_{k^\prime}^{+}a^{-}_{k^\prime}}  +\cdots,
\label{traceK0}
\end{equation}
where the ellipsis stands for antisymmetric terms, which do not affect the form of the gap equation.
Indeed, when contracted with the gluon propagator, which is symmetric in Lorentz indices, all 
antisymmetric terms will vanish.

At asymptotic densities, we can also neglect all corrections due to nonzero $m$ and $ \delta\mu $. By taking into 
account that the main contribution to the momentum integral on the right-hand side of the gap equation comes from the  
vicinity of the Fermi surface ($k^\prime\simeq k_F=\sqrt{\bar{\mu}^2-m^2}$), we can make the following approximation
for the trace:
\begin{equation}
\mbox{tr}\left[ \gamma^{\mu}K^{(0)} (\omega^\prime,\bm{k}^\prime) \gamma^{\nu}\right] \simeq \frac{2 g^{\mu\nu} }{a^{-}_{k^\prime}} .
\end{equation}
Note that, in the vicinity of the Fermi surface, one has  
\begin{eqnarray}
  a^{-}_{k^\prime} & = &  (\omega_E^\prime)^2 +\xi_{k^\prime}^2+\Delta^2 \ll \bar{\mu}^2 , \\
  a^{+}_{k^\prime} & = & 4\bar{\mu}^2 +4\bar{\mu}\xi_{k^\prime} +a^{-}_{k^\prime} \simeq 4\bar{\mu} (\bar{\mu}+\xi_{k^\prime}),
\end{eqnarray}
where $\xi_{k^\prime} \equiv E_{k^\prime}-\mu \simeq k^\prime-k_F$. 

The resulting equation coincides with the known form of the gap equation in the case of zero magnetic field 
studied in Refs.~\cite{Son1999, Schafer1999, Shovkovy1999, Hong:1999ru, Hong2000, Hsu2000, Pisarski2000}. 
In our notation, the corresponding solution for the gap function reads 
\begin{equation}
 |\Delta^{(0)}| \simeq  \Lambda \exp(-\frac{3\pi^2}{\sqrt{2}g}+1),
\end{equation}
where $\Lambda={4(2\mu)^3}/({\pi m_D^2})$. 

In order to find the correction to the gap function due to nonzero magnetic field, let us include the 
approximate kernel up to second order in the magnetic field. After taking traces on the both sides 
of the equation, we obtain
\begin{eqnarray}
\Delta^{(B)}(\omega_E) & = & \frac{g^2}{6}\int\frac{d\omega_E^\prime}{2\pi}
   \int \frac{d^3\bm{k}^\prime}{(2\pi)^3}   \Delta^{(B)}(\omega_E^\prime) 
   \mbox{tr}\left[ \gamma^{\mu}K^{(0)}(\omega^\prime,\bm{k}^\prime)
   \gamma^{\nu}\right] D_{\mu\nu}(\omega-\omega^\prime,\bm{k}-\bm{k}^\prime)\nonumber\\
   & + & \frac{g^2}{6}\int\frac{d\omega_E^\prime}{2\pi}
   \int \frac{d^3\bm{k}^\prime}{(2\pi)^3}   \Delta^{(B)}(\omega_E^\prime) 
   \mbox{tr}\left[ \gamma^{\mu}K^{(2)}(\omega^\prime,\bm{k}^\prime)
   \gamma^{\nu}\right] D_{\mu\nu}(\omega-\omega^\prime,\bm{k}-\bm{k}^\prime). 
\label{GapEq2nd}
\end{eqnarray}
In addition to the result in Eq.~(\ref{traceK0}), this equation also contains the trace of the second order correction
to the kernel. The corresponding approximate expression in the vicinity of the Fermi surface reads
\begin{equation}
\mbox{tr}\left[ \gamma^{\mu}K^{(2)} (\omega^\prime,\bm{k}^\prime)\gamma^{\nu}\right]  \simeq g^{\mu\nu} 
\frac{N_{k^\prime} (k_\perp^\prime)^2 }{2\bar{\mu}^4(a^{-}_{k^\prime})^4}  (\tilde{e}\tilde{Q}\tilde{B})^2 +\cdots,
\end{equation}
where $N_{k^\prime}\simeq 4 \bar{\mu}\xi_{k^\prime} (2\xi_{k^\prime}^2-a^{-}_{k^\prime})-24\xi_{k^\prime}^4
+16 a^{-}_{k^\prime}\xi_{k^\prime}^2-(a^{-}_{k^\prime})^2$ and the ellipsis denotes antisymmetric terms.
Let us point out that the only directional dependence of this trace comes through the overall factor $(k^\prime_\perp)^2 
\equiv (k^\prime)^2(1-\cos^2\theta_{Bk^\prime})$, where $\theta_{Bk^\prime}$ denotes the angle between the 
direction of the magnetic $\bm{B}$ and the momentum $\bm{k}^\prime$. (Strictly speaking, in a self-consistent
analysis, the gap function on the right-hand side will also have a directional dependence and will affect 
the angular integration. The corresponding effects are expected to be very small and will be neglected 
in the simplified analysis here.) The integrand on the right-hand
side of Eq.~(\ref{GapEq2nd}) has an additional directional dependence in the gluon propagator [see 
Eq.~(\ref{gluon M})], which is a function of the polar angle $\theta \equiv \theta_{kk^\prime}$ (i.e., the 
polar angular coordinate of vector 
$\bm{k}^\prime$ measured from the direction of the external vector $\bm{k}$). With this convention for 
angular coordinates, it is convenient to use the following relation:
\begin{equation}
\cos\theta_{Bk^\prime} = \sin\theta \, \sin\theta_{Bk} \,\cos(\phi-\phi_{Bk} )+\cos\theta \,\cos\theta_{Bk},
\end{equation}
in order to rewrite the expression for $(k^\prime_\perp)^2$ in terms of the angular integration variables  
$\theta$ (polar angle) and $\phi$ (azimuthal angle). Now we can easily perform the angular integration 
on the right-hand side of the gap equation. The results for the two types of angular integrations, namely
with the electric and magnetic part of the gluon propagator, read
\begin{eqnarray}
   A_{\rm el}&=&\int \frac{(1-\cos^2\theta_{Bk^\prime}) \sin\theta \,  d\theta d\phi }
   {M^2-2k^\prime k\cos\theta}
 =   \frac{\pi }{8 (k^\prime)^3 k^3}\Bigg[2k^\prime k M^2 \left[1+3\cos(2\theta_{Bk}) \right]\nonumber \\
   &+&\frac{1}{2} \left(4 (k^\prime)^2 k^2 \left[3+\cos(2\theta_{Bk})\right] -M^4 \left[1+3\cos(2\theta_{Bk})\right]\right)
   \ln\frac{M^2+2k^\prime k}{M^2-2k^\prime k} \Bigg],\\
   A_{\rm mag}&=& \int (1-\cos^2\theta_{Bk^\prime}) \sin\theta \,  d\theta d\phi  \frac{2\left[(k^\prime)^2+k^2-2k^\prime k\cos\theta \right]^{1/2}}
   {\left[(k^\prime)^2+k^2-2k^\prime k\cos\theta \right]^{3/2}+\omega_l^3} 
   \nonumber \\
   &=&\frac{2\pi}{3k^\prime k}\left[1+\cos^2\theta_{Bk}+\left(\frac{(k^\prime)^2+k^2}{2k^\prime k}\right)^2
   \left(1-3\cos^2\theta_{Bk}\right)\right]\ln\frac{(k^\prime + k)^3+\omega_l^3}{|k^\prime - k|^3+\omega_l^3} \nonumber \\
   &+&\frac{\pi \omega_l^2}{2(k^\prime k)^3} \left(1-3\cos^2\theta_{Bk}\right) \left[
   \omega_l^2 \int_{x_{\rm min}}^{x_{\rm max}}\frac{x^6 dx}{x^3+1}
   -2\left[(k^\prime)^2+k^2\right] \int_{x_{\rm min}}^{x_{\rm max}}\frac{x^4 dx}{x^3+1}
   \right],
   \end{eqnarray}
where $M^2=(\omega_E^\prime-\omega_E)^2 +(k^\prime)^2+k^2+m_D^2$. In order to simplify 
the calculation of $A_{\rm mag}$, it is convenient to change the integration variable $\theta$ to the 
new dimensionless variable $x=(1/\omega_l)\sqrt{(k^\prime)^2+k^2-2k^\prime k\cos\theta}$. 
Note that $\sin\theta d\theta =\omega_l^2 x dx/(k^\prime k)$ and the new range of integration is from 
$x_{\rm min}=|k^\prime - k|/\omega_l$ to $x_{\rm max}=(k^\prime + k)/\omega_l$.

In the vicinity of the Fermi surface, the approximate results for these integrals read
\begin{eqnarray}
   A_{\rm el}&\simeq &\frac{\pi \sin^2\theta_{Bk} }{\bar{\mu}^2} 
   \ln\frac{(2\bar{\mu})^2}{(\omega_E^\prime-\omega_E)^2 +(k^\prime-k)^2+m_D^2} +\cdots , \\
   A_{\rm mag}&\simeq & \frac{4\pi \sin^2\theta_{Bk} }{3\bar{\mu}^2}\ln\frac{(2\bar{\mu})^3}{|k^\prime - k|^3+\omega_l^3}+\cdots,
\end{eqnarray} 
where the ellipses denote the subleading terms.

By making use of the above intermediate results, we arrive at the following form of the gap equation,
\begin{eqnarray}
  \Delta^{(B)}(\omega_E)
    &=& \frac{2g^2}{9}  \int_{-\infty}^{\infty} \frac{d\omega_E^\prime }{(2\pi)} \int \frac{d \xi_{k^\prime}}{(2\pi)^2} 
   \frac{\Delta^{(B)}(\omega_E^\prime)}{a^{-}_{k^\prime}}   \ln\frac{(2\bar{\mu})^3}{|k^\prime - k|^3+\omega_l^3} 
   \nonumber\\
  &\times&  \left(1+ \frac{\left[-24\xi_{k^\prime}^4
+16 a^{-}_{k^\prime}\xi_{k^\prime}^2-(a^{-}_{k^\prime})^2\right] (\tilde{e}\tilde{Q}\tilde{B})^2}{(2\bar{\mu})^2 (a^{-}_{k^\prime})^3} 
    \sin^2\theta_{Bk} \right) .
\end{eqnarray}
Recall that $\omega_l^3=(\pi/4)m_D^2|\omega_E^\prime-\omega_E|$. Integrating over the momentum,
we arrive at 
\begin{eqnarray}
  \Delta^{(B)}(\omega_E)
    &=& \frac{g^2}{36\pi^2} \int_{-\infty}^{\infty} d\omega_E^\prime  
   \Delta^{(B)}(\omega_E^\prime)\Bigg[\frac{1}{\sqrt{(\omega_E^\prime)^2+(\Delta^{(B)})^2}}
   \ln\frac{\Lambda}{|\omega_E^\prime-\omega_E|}+\nonumber\\
  &&+\frac{9\omega_l^{15} (\tilde{e}\tilde{Q}\tilde{B})^2  \sin^2\theta_{Bk} }{4\bar{\mu}^2\left(\omega_l^6+\left[(\omega_E^\prime)^2+(\Delta^{(B)})^2\right]^3\right)^3}\ln\frac{\omega_l}{|\omega_E^\prime-\omega_E|}\Bigg].
\end{eqnarray}
To get a rough estimate, let us take an infrared cutoff in the energy integration at $\omega^\prime_{\rm IR}\simeq\Delta^{(B)}$
and drop the dependence on $\Delta^{(B)}$ in the denominator of the integrand. Then, we have 
\begin{eqnarray}
  \Delta^{(B)}
    &\simeq & \frac{g^2}{18\pi^2}  \int_{\Delta^{(B)}}^{\Lambda} d\omega_E^\prime  \frac{\Delta^{(B)}}{|\omega_E^\prime|}
   \left(1+\frac{54 (\tilde{e}\tilde{Q}\tilde{B})^2  \sin^2\theta_{Bk} }{\pi \bar{\mu}^2 m_D^2}\right)
    \ln\frac{\Lambda}{|\omega_E^\prime|}.
\end{eqnarray}
This means that the magnetic field correction is equivalent to an effective increase of the coupling constant,
i.e.,
\begin{equation}
g^2\to g_{\rm eff}^2= g^2\left(1+\frac{54 \pi (\tilde{e}\tilde{Q}\tilde{B})^2 }{g^2 \bar{\mu}^4} \sin^2\theta_{Bk} \right),
\end{equation}
where we used the definition of the Debye mass $m_D^2=(g\bar{\mu}/\pi)^2$.

\end{document}